\documentclass{elsart}
\pdfoutput=1
\usepackage[pdftex]{graphicx}
\usepackage[unicode=true, bookmarks=true,bookmarksnumbered=true,bookmarksopen=false, breaklinks=false,backref=false,pagebackref=false, colorlinks=true]{hyperref}
\hypersetup{pdftitle={Time dependent simulation of the Driven Lid Cavity at High Reynolds Number},pdfauthor={Nuno Cardoso, Pedro Bicudo},linkcolor=black,citecolor=black, urlcolor=black, filecolor=black,pdfpagelayout=OneColumn,pdfnewwindow=true,pdfstartview=XYZ, plainpages=false}

\usepackage{amssymb}
\usepackage{amsmath}
\usepackage{subfig}
\usepackage{float}
\usepackage{url}
\usepackage[T1]{fontenc}
\usepackage[latin1]{inputenc}
\usepackage[english]{babel}

\newcommand{\re}{\operatorname{Re}}

\begin{document}
\begin{frontmatter}

\title{Time dependent simulation of the Driven Lid Cavity at High Reynolds
Number}

\author{N. Cardoso and P. Bicudo}

\address{CFTP, Departamento de Física, Instituto Superior Técnico, Av. Rovisco
Pais, 1049-001 Lisboa, Portugal}
\begin{abstract}
In this work, numerical solutions of the two dimensional time dependent
incompressible flow, in a driven cavity at high Reynolds number $\re$,
are presented. At high $\re$, there is a controversy. Some studies
predicted that the flow is steady, others found time dependent non-steady
flow, either periodic or aperiodic. In this study, the driven lid
cavity is successfully solved using a very fine grid mesh, for $\re$
up to $30\,000$. We discretize the Vorticity-Stream formulation of
the Navier-Stokes equation with the SSPRK(5,4) scheme in a $1024\times1024$
grid. Using this very fine grid, the results obtained converge to
a stationary solution. Detailed results for $\re$ between $5\,000$
and $30\,000$ are presented. The driven lid cavity problem is solved
with a NVIDIA GPU using the CUDA programming environment with double
precision.\end{abstract}
\end{frontmatter}

\section{Introduction}

The driven lid cavity flow is one of the most studied problems of
the physics of fluids. The simple geometry makes the problem easy
to solve numerically and apply boundary conditions. Despite being
a problem rather studied, there are still some questions and controversy
about what happens at high Reynolds numbers.

In the literature it is possible to find numerous studies about the
driven cavity flow, however the nature of the flow at high Reynolds
number is still not agreed upon. Driven cavity flow serve as a benchmark
problem for numerical methods in terms of numerical efficiency and
accuracy. Erturk \cite{Erturk08} grouped these numerical studies
into three categories. In the first category are the studies with
numerical solutions of 2-D steady incompressible flow at high Reynolds
numbers. Some of these studies are Erturk et. al. \cite{Erturk05},
Erturk and Gokcol \cite{Erturk06}, Barragy and Carey \cite{Barragy97},
Schreiber and Keller \cite{Schreiber83}, Benjamin and Denny \cite{Benjamin79},
Liao and Zhu \cite{Liao96}, Ghia et. al. \cite{Ghia82} for $\re=10\,000$.
Barragy and Carey \cite{Barragy97} have also presented solutions
for $\re=12\,500$. Erturk et. al. \cite{Erturk05} and also Erturk
and Gokcol \cite{Erturk06} have presented steady solutions up to
$\re=30\,000$. In the second category we have the studies about hydrodynamic
stability analysis, Fortin et. al. \cite{Fortin97}, Gervais et. al.
\cite{Gervais97}, Sahin and Owens \cite{Sahin03} and Abouhamza and
Pierre \cite{Abouhamza03}. And the third category includes the studies
about the steady to unsteady transition flow through a Direct Numerical
Simulation (DNS) and the transition Reynolds number. Some of the studies
found in the literature are Auteri, Parolini and Quartepelle \cite{Auteri02},
Peng, Shiau and Hwang \cite{Peng03}, Tiesinga, Wubs and Veldman \cite{Tiesinga02},
Poliashenko and Aidun \cite{Poliashenko95}, Cazemier, Verstappen
and Veldman \cite{Cazemier98}, Goyon \cite{Goyon96}, Wan, Zhou and
Wei \cite{Wan02} and Liffman \cite{Liffman96}.

Erturk \cite{Erturk04} studied the numerical solutions of 2-D steady
flow, with time independent equations, for $\re\leq20\,000$ and found
that at high Reynolds numbers the solutions obtained depend of the
mesh spacing. For coarse meshes, at high Reynolds numbers, he found
no convergence. But for a fine grid he finds a steady solution. Thus
he claimed that the true solution is steady and to be found numerically
a very fine grid is needed.

Zhen-Hua et al. \cite{Zhen06} with multi-relaxation-time (MRT model)
lattice Boltzmann method obtained solutions for up to $\re=1\,000\,000$,
but the grid used ($256\times256$) in the MRT model are coarse compared
to the article of Erturk, and the solutions obtained by them are not
steady.

Notice that the existence of solutions to the stationary flow equation,
does not necessarily imply that the natural solution is stationary.
Let us illustrate this claim with the well known case of metastable
water. At the normal pressure and at temperatures below freezing,
two phases of water, i.e., two solutions of the non-linear equations
for the water, exist. The stable solution is composed of solid water
(ice) and the metastable is composed of supercooled water. In the
same way the existence of a stationary solution to the flow equations,
does not necessarily exclude the existence of a second, unsteady,
solution. The solution of the time-dependent equation is necessary
to discriminate which is the true solution, occurring naturally in
the lab.

Since different authors find different solutions, in this paper we
examine if for high Reynolds numbers as high as $\re=30\,000$ the
numerical solutions of 2-D time dependent incompressible flow in a
driven cavity become steady after a given time. Importantly, our equations
are time-dependent, and our grid is very fine, thus we expect to obtain
the true solution. We use the vorticity-stream function formulation
to the Navier-Stokes equation.

This study is divided in five sections. The first section, the objective
is presented and the existing works on the subject. In the second
section, we present the theory and the fundamental equations used
in the numeric calculations. In the third section, we describe the
driven lid cavity problem, numeric equations, boundary conditions
and the algorithm used. In the section four, we present the results
and a comparison with existing studies about the cavity flow. Finally,
in the section five, we present the conclusions about this study.

\section{Vorticity-Stream Function Formulation}

The fundamental equations of fluid dynamics are based on three universal
laws of conservation: mass, moment and energy conservation. In general,
there are two ways to solve numerically the Navier-Stokes equations.
The first is proceeding with primitive variables formulation, $u$,
$v$ e $p$. The second is using the vorticity-stream function approach.

Using the stream function, $\psi$ and the vorticity, $\omega$, in
place of the primitive variables $u$, $v$ e $p$, where these quantities
are defined by: \begin{equation}
\omega=\nabla\times\overrightarrow{\mathbf{u}}=\frac{\partial v}{\partial x}-\frac{\partial u}{\partial y}\label{vort0}\end{equation}
 and \begin{equation}
\left\{ \begin{array}{l}
\frac{\partial\psi}{\partial y}=u\\
\frac{\partial\psi}{\partial x}=-v\end{array}\right.\label{stream0}\end{equation}

Combining the equations (\ref{vort0}) and (\ref{stream0}), \begin{equation}
\Delta\psi=\nabla^{2}\psi=\frac{\partial^{2}\psi}{\partial x^{2}}+\frac{\partial^{2}\psi}{\partial y^{2}}=-\omega\label{str}\end{equation}

The main reason for enter the stream function is that, for runoffs
in which $\rho$ and $\nu$ are constants, the equation of continuity
is satisfied.

The vorticity-stream function is given by\begin{align}
 & \frac{\partial\omega}{\partial t}=-\frac{\partial\psi}{\partial y}\frac{\partial\omega}{\partial x}+\frac{\partial\psi}{\partial x}\frac{\partial\omega}{\partial y}+\nu\left[\frac{\partial^{2}\omega}{\partial x^{2}}+\frac{\partial^{2}\omega}{\partial y^{2}}\right]\label{eq:vortfinal}\\
 & \frac{\partial^{2}\psi}{\partial x^{2}}+\frac{\partial^{2}\psi}{\partial y^{2}}=-\omega\end{align}
where $\overrightarrow{u}=\left(u,v\right)$ is the fluid velocity,
$p$ the pressure, $\rho$ the fluid density and $\nu$ kinematic
viscosity. The kinematic viscosity is given by:\begin{equation}
\nu=\frac{\eta}{\rho}\end{equation}
 where $\eta$ is the viscosity.

\section{Numerical Method}

\begin{figure}[H]
\begin{centering}
\includegraphics[width=7cm]{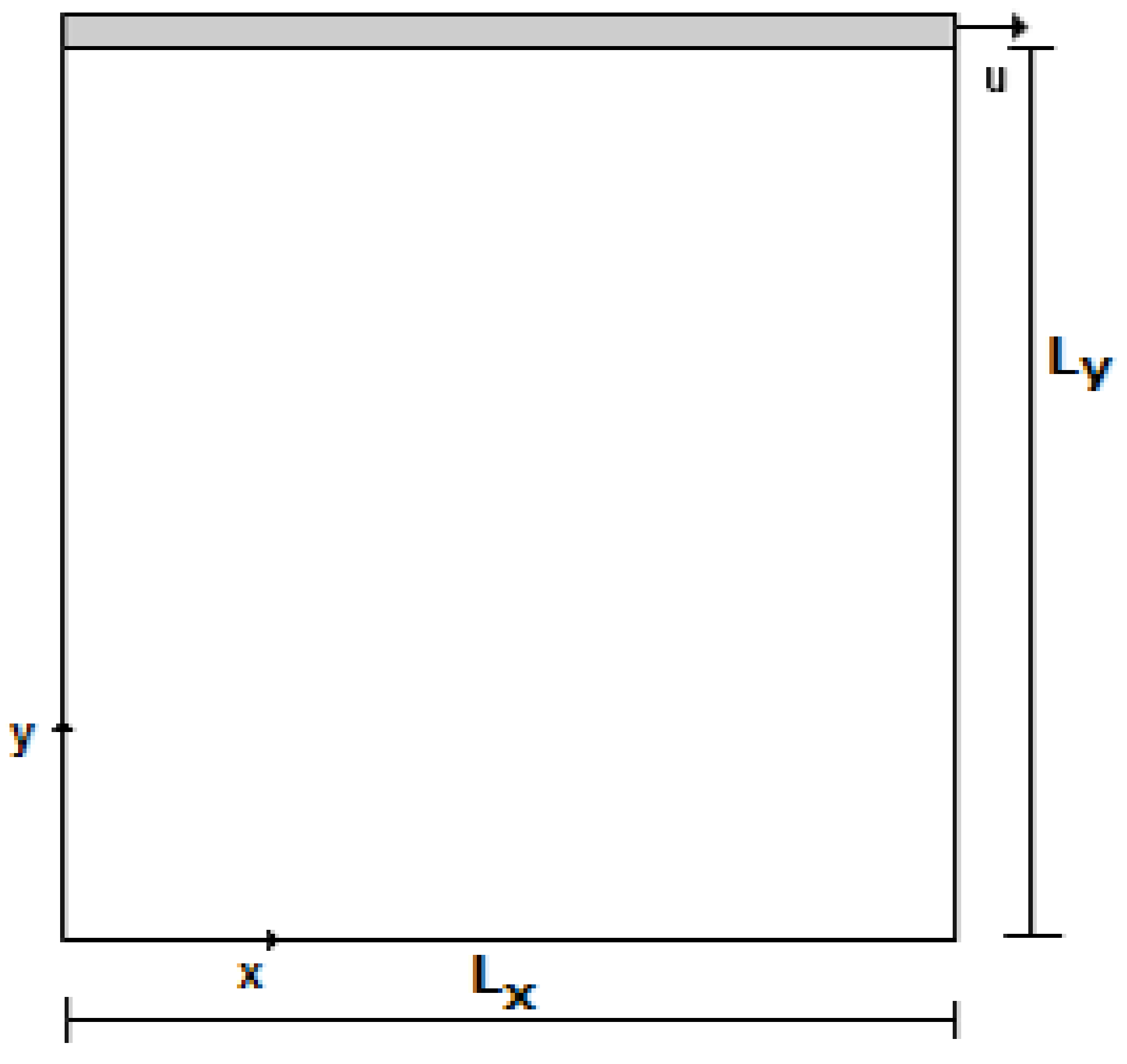}
\par\end{centering}

\caption{Driven Lid Cavity.\label{figdlc} }

\end{figure}

In this section we present the driven lid cavity problem, the numerical
equations, the boundary conditions and the algorithm. In the figure
\ref{figdlc} is represented the outline problem to study. The top
of the wall cavity moves with speed $u=U$.

Numeric vorticity equation is given by: \begin{eqnarray}
\omega_{i,j}^{n+1} & = & \omega_{i,j}^{n}+\Delta tL\left(\omega^{n}\right)\label{eq:vortnum}\end{eqnarray}
with\begin{eqnarray}
L\left(\omega^{n}\right) & = & -u_{i,j}\frac{\omega_{i+1,j}^{n}-\omega_{i-1,j}^{n}}{2\Delta x}-v_{i,j}\frac{\omega_{i,j+1}^{n}-\omega_{i,j-1}^{n}}{2\Delta y}+\nonumber \\
 &  & +\nu\left(\frac{\omega_{i+1,j}^{n}-2\omega_{i,j}^{n}+\omega_{i-1,j}^{n}}{\left(\Delta x\right)^{2}}+\frac{\omega_{i,j+1}^{n}-2\omega_{i,j}^{n}+\omega_{i,j-1}^{n}}{\left(\Delta y\right)^{2}}\right)\label{eq:Lwn}\end{eqnarray}

The five stage fourth order SSPRK developed by Ruuth and Spiteri,
\cite{Spiteri2002} , and guaranteed optimal \cite{Ruuth2006}, is
given by:

\begin{align}
\omega^{(1)} & =\omega^{n}+0.39175222657\,\Delta t\, L\left(\omega^{n}\right)\label{eq:SSPRK54_1}\\
\omega^{(2)} & =0.444370493651235\,\omega^{n}+0.555626506348765\,\omega^{(1)}\label{eq:SSPRK54_2}\\
 & +0.368410593050371\,\Delta t\, L\left(\omega^{(1)}\right)\nonumber \\
\omega^{(3)} & =0.620101851488403\,\omega^{n}+0.379898148511597\,\omega^{(2)}\label{eq:SSPRK54_3}\\
 & +0.251891774271694\,\Delta t\, L\left(\omega^{(2)}\right)\nonumber \\
\omega^{(4)} & =0.178079954393132\,\omega^{n}+0.821920045606868\,\omega^{(3)}\label{eq:SSPRK54_4}\\
 & +0.544974750228521\,\Delta t\, L\left(\omega^{(3)}\right)\nonumber \\
\omega^{n+1} & =0.517231671970585\,\omega^{(2)}\label{eq:SSPRK54_5}\\
 & +0.096059710526147\,\omega^{(3)}+0.063692468666290\,\Delta t\, L\left(\omega^{(3)}\right)\nonumber \\
 & +0.386708617503269\,\omega^{(4)}+0.226007483236906\,\Delta t\, L\left(\omega^{(4)}\right)\nonumber \end{align}

To get the stream function, we apply the fourth order method, \cite{Erturk06},
to solve the Poisson equation (\ref{str}), \begin{equation}
\frac{\partial^{2}\psi}{\partial x^{2}}+\frac{\partial^{2}\psi}{\partial y^{2}}=-\omega-\frac{\Delta x^{2}}{12}\frac{\partial^{2}\omega}{\partial x^{2}}-\frac{\Delta y^{2}}{12}\frac{\partial^{2}\omega}{\partial y^{2}}-\left(\frac{\Delta x^{2}}{12}+\frac{\Delta y^{2}}{12}\right)\frac{\partial^{4}\psi}{\partial x^{2}\partial y^{2}}\label{eq:stream_4O}\end{equation}

To solve the fourth order Stream equation, we use the Successive Over
Relaxation method (SOR), since this method converge much faster than
the traditional methods (methods of Jacobi and Gauss-Seidel). The
idea is to use the available current estimates from other locations
($\psi_{i+1,j}^{n+1}$) when they become available and use the estimates
of current positions ($\psi_{i,j}^{n}$) to improve the estimative
of $\psi_{i,j}^{n+1}$: \begin{equation}
\psi^{n+1}=\frac{3}{5}\frac{\beta}{a}\left(A+B+C\right)+\left(1-\beta\right)\psi_{i,j}^{n}\label{SequationFinal}\end{equation}
where $\beta$ is the relaxation parameter, and

\begin{eqnarray}
A & = & b\left(\psi_{i+1j}^{n}+\psi_{i-1j}^{n+1}\right)+c\left(\psi_{ij+1}^{n}+\psi_{ij-1}^{n+1}\right)\label{eq:-3}\\
B & = & \frac{1}{12}\left(\omega_{i+1j}^{n}+\omega_{ij+1}^{n}+8\omega_{ij}^{n}+\omega_{i-1j}^{n}+\omega_{ij-1}^{n}\right)\label{eq:-4}\\
C & = & \frac{1}{12}a\left(\psi_{i+1j+1}^{n}-2\psi_{ij+1}^{n}+\psi_{i-1j+1}^{n+1}-2\psi_{i+1j}^{n}-\right.\label{eq:-5}\\
 &  & \left.-2\psi_{i-1j}^{n+1}+\psi_{i+1j-1}^{n}-2\psi_{ij-1}^{n+1}+\psi_{i-1j-1}^{n+1}\right)\nonumber \\
\nonumber \\a & = & b+c\label{eq:-6}\\
b & = & \frac{1}{\Delta x^{2}}\label{eq:-7}\\
c & = & \frac{1}{\Delta y^{2}}\label{eq:-8}\end{eqnarray}

To obtain the velocity ($u$ e $v$) it is only need to numerically
solve the equation (\ref{stream0}), \begin{equation}
\left\{ \begin{array}{l}
u_{i,j}=\frac{\psi_{i,j+1}-\psi_{i,j-1}}{2\Delta y}\\
v_{i,j}=-\frac{\psi_{i+1,j}-\psi_{i-1,j}}{2\Delta x}\end{array}\right.\label{vel}\end{equation}

\subsection{Boundary Conditions}

The boundary conditions in the four walls are given by:
\begin{itemize}
\item left and right walls: \begin{equation}
\left\{ \begin{array}{l}
u=0\\
v=-\frac{\partial\psi}{\partial x}=U_{wall}\\
S=0\\
\omega=-\frac{\partial^{2}\psi}{\partial x^{2}}\end{array}\right.\end{equation}

\item top and bottom walls: \begin{equation}
\left\{ \begin{array}{l}
u=\frac{\partial\psi}{\partial y}=U_{wall}\\
v=0\\
S=0\\
\omega=-\frac{\partial^{2}\psi}{\partial y^{2}}\end{array}\right.\end{equation}

\end{itemize}
where $U_{wall}$ is the uniform velocity for the translation of the
top wall and zero for the other three walls.

In order to solve equation (\ref{eq:SSPRK54_5}), Thom\textquoteright{}s
method is used for calculating vorticity on the boundaries, therefore
the boundary conditions are given by \begin{equation}
\left\{ \begin{array}{l}
\omega_{i,1}=-2\frac{\psi_{i,2}}{\Delta x^{2}}\\
\omega_{i,n_{y}}=-2\frac{\psi_{i,n_{y}-1}}{\Delta x^{2}}-2\frac{U}{\Delta x}\\
\omega_{1,j}=-2\frac{\psi_{2,j}}{\Delta y^{2}}\\
\omega_{n_{x},j}=-2\frac{\psi_{n_{x}-1,j}}{\Delta y^{2}}\end{array}\right.\end{equation}

\subsection{Algorithm}

\begin{figure}[H]
\begin{centering}
\includegraphics[width=7cm]{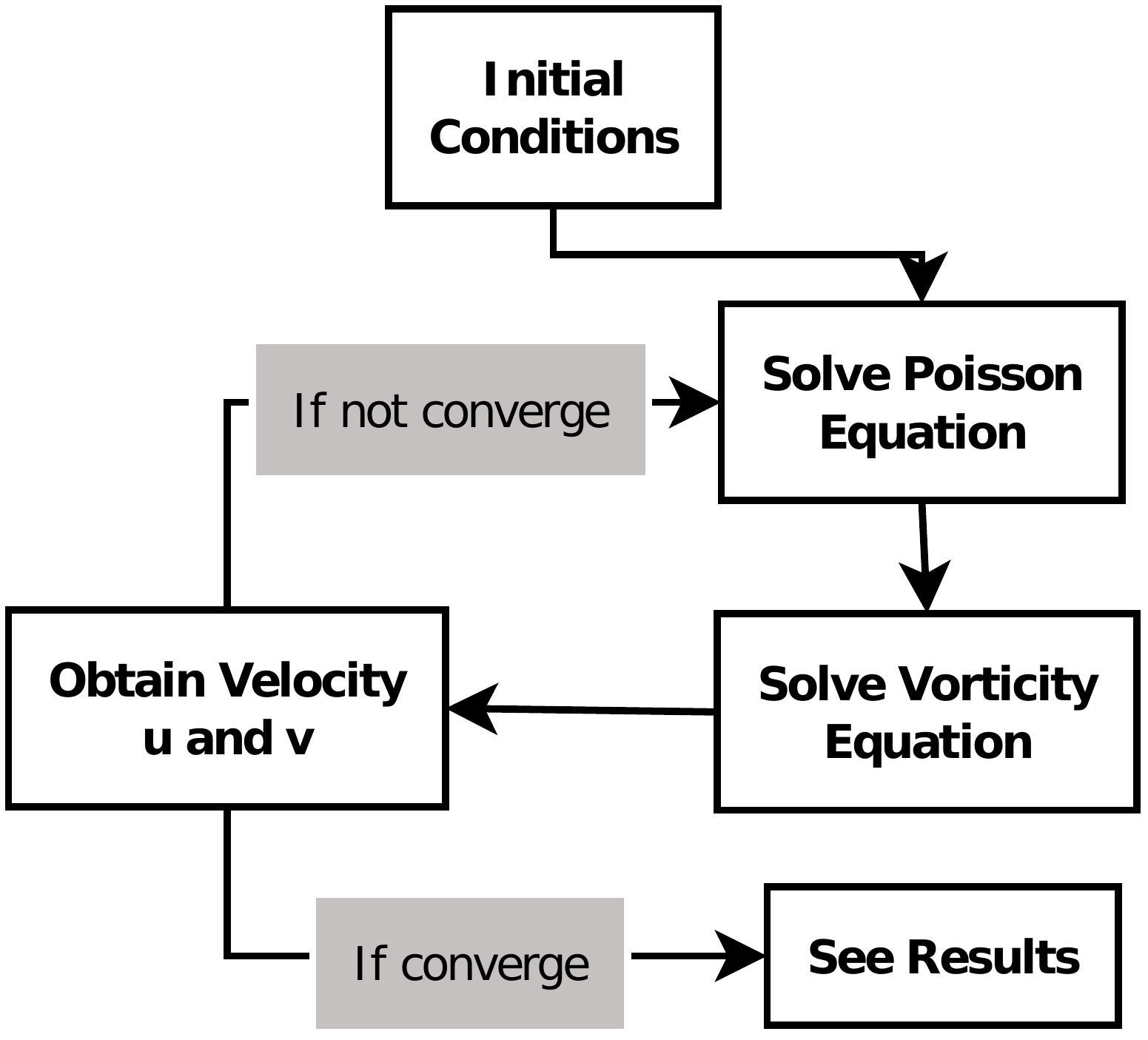}
\par\end{centering}

\caption{Scheme for solving the problem of the driven lid cavity using the
vorticity-stream function approach.}

\centering{}\label{algsv}
\end{figure}

An algorithm for solving the problem of the driven lid cavity using
the vorticity-stream function approach is given by the scheme in figure
\ref{algsv}.

To ensure the stability and convergence of the algorithm, $\Delta t$
should be small enough to a given viscosity, $\nu$, and resolution
of the grid, $\Delta x\Delta y$. The Reynolds number, $\re$, can
be calculated using the kinematic viscosity, $\nu$, and the conditions
of the cavity: \[
\begin{array}{l}
\re=\frac{U_{wall}L_{y}}{\nu}=\frac{\rho U_{wall}L_{y}}{\eta}\\
L=L_{x}L_{y}\end{array}\]
 in the following calculations it was considered: \begin{align*}
\rho & =1Kg/m^{3}\\
L & =L_{x}=L_{y}=1\\
U_{wall} & =1\\
N & =n_{x}=n_{y}\end{align*}
 Using this values, it's the same using dimensionless variables.

In this study, we defined $RES$ by the following equations

\begin{eqnarray}
RES_{\psi} & = & \frac{1}{N}\sum_{i,j}\left\vert \psi_{i,j}^{n+1}-\psi_{i,j}^{n}\right\vert \label{eq:resS}\\
RES_{\omega} & = & \frac{1}{N}\sum_{i,j}\left\vert \omega_{i,j}^{n+1}-\omega_{i,j}^{n}\right\vert \label{eq:resV}\end{eqnarray}
for stream function and vorticity respectively, as convergence criteria
to the steady state.

\section{Results}

In this work, we present results of the cavity flow from $\re=5\,000$
up to $\re=30\,000$.

\begin{figure}
\begin{centering}
\includegraphics[width=8cm]{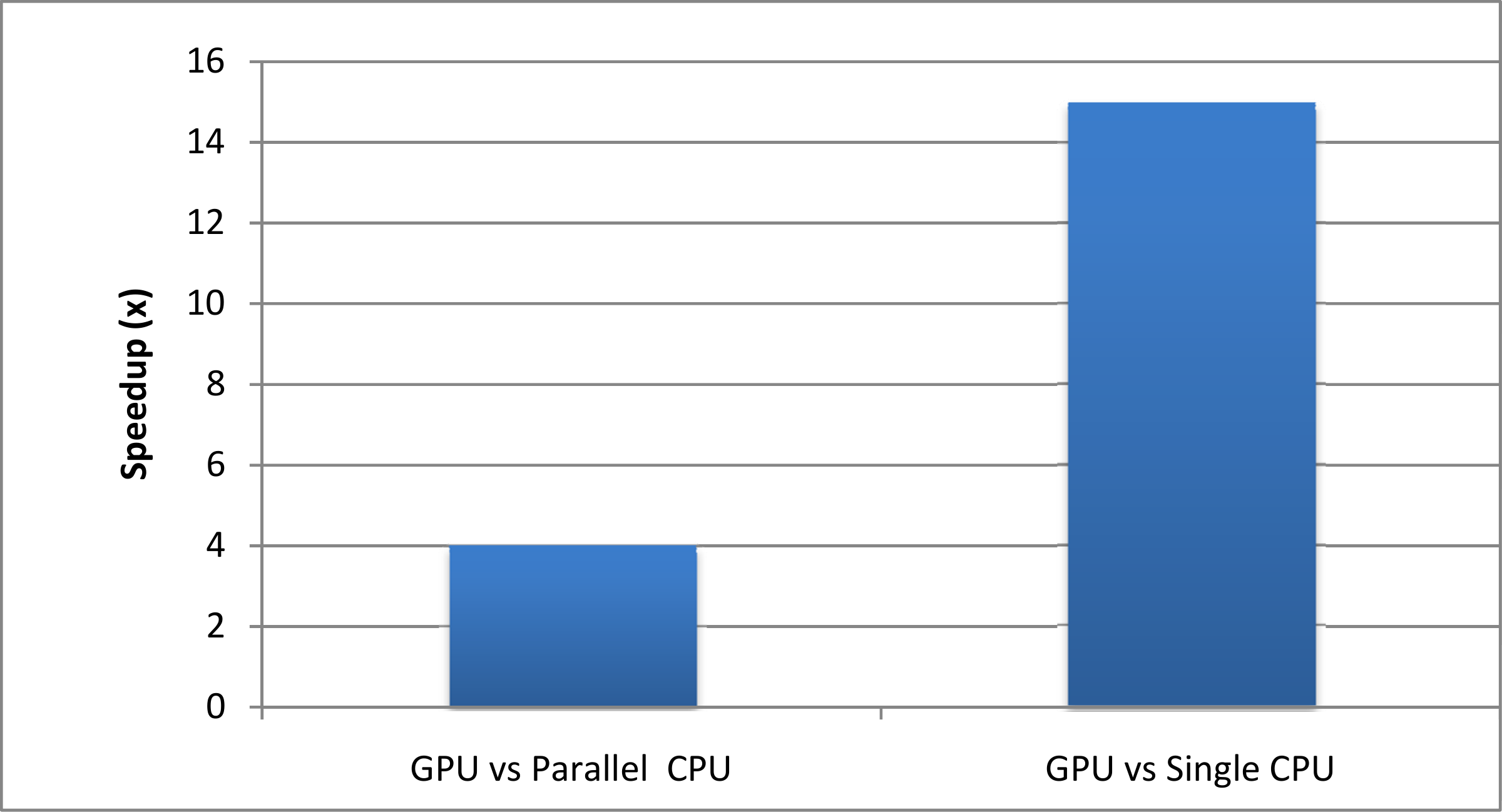}
\par\end{centering}

\caption{GPU code speedup relative to the serial and parallel CPU code for
a grid of 1024$\times$1024.\label{fig:GPU_CPU}}

\end{figure}

The numerical code was written for use in CPU's (Intel(R) Core(TM)2
Quad CPU Q9450 @ 2.66GHz) and in GPU's (NVIDIA GEFORCE 280 GTX GPU
with double precision capabilities). For the CPU, we use OPENMP with
C language and for GPU we use CUDA language. Most of the GPU's only
support single precision but the recent GPU's have included support
for double precision, although using double precision is almost eight
times slower than single precision. Nevertheless, the GPU used is
faster than the most recent Intel Quad core. Figure \ref{fig:GPU_CPU}
summarizes our GPU code performance relative to the serial CPU version
of our code and CPU parallel code performance relative to the serial
CPU version of our code using double precision in all cases. In the
parallel CPU version code we tested it in a Quad Core CPU with OPENMP.

In order to obtain a good convergence of the method we need to choose
an appropriate value for the $\beta$ parameter, the relaxation parameter,
in equation (\ref{SequationFinal}). We make several tests in order
to see which value is the best and this parameter varies from 0.8
to 0.1 as the Reynolds number increase.

Using the algorithm described above, we started with a grid of $220\times220$
points for $\re=5\,000$ and $\re=10\,000$ and the solution converges
to a steady state in time.

For $\re=15\,000$ we obtain a solution that converges to a steady
state with a grid of $300\times300$. With this small lattice we cannot
obtain a solution for the Reynolds number above $15\,000$. Erturk
et. al. \cite{Erturk05} said that they have to use a grid of $1025\times1025$
for $\re>15\,000$. So, when we use a grid of $1024\times1024$ for
$\re\geq15\,000$, we obtain a steady solution up to $\re=30\,000$.
The choice of this grid is due to the GPU architecture. Table \ref{tab:grid}
shows the grid mesh used at various Reynolds numbers for the final
results. For $\re$ up to $20\,000$, the algorithm ended when the
value obtained by $RES_{\psi}$ and $RES_{\omega}$ were less than
$10^{-12}$ and $10^{-10}$ for stream function and vorticity respectively.
For $\re=25\,000$ and $\re=30\,000$, the values considered for $RES_{\psi}$ and $RES_{\omega}$
were $10^{-10}$ and $10^{-8}$. Such values are more than satisfactory,
demonstrating that the solution converges to a steady state.

\begin{table}
\begin{centering}
\begin{tabular}{|c|c|c|c|}
\hline
$\re$  & Grid  & $\Delta t\ (s)$  & Figures\tabularnewline
\hline
\hline
$5\,000$  & $1024\times1024$  & $0.001$  & \ref{fig:5kS} and \ref{fig:5kV}\tabularnewline
\hline
$10\,000$  & $1024\times1024$ & $0.001$  & \ref{fig:10kS} and \ref{fig:10kV}\tabularnewline
\hline
$15\,000$  & $1024\times1024$ & $0.001$  & \ref{fig:15kS} and \ref{fig:15kV}\tabularnewline
\hline
$20\,000$  & $1024\times1024$ & $0.001$  & \ref{fig:20kS} and \ref{fig:20kV}\tabularnewline
\hline
$25\,000$  & $1024\times1024$ & $0.001$  & \ref{fig:25kS} and \ref{fig:25kV}\tabularnewline
\hline
$30\,000$  & $1024\times1024$ & $0.001$  & \ref{fig:30kS} and \ref{fig:30kV}\tabularnewline
\hline
\end{tabular}
\par\end{centering}

\caption{Grid mesh.}

\label{tab:grid}
\end{table}

\begin{figure}
\begin{centering}
\includegraphics[width=8cm]{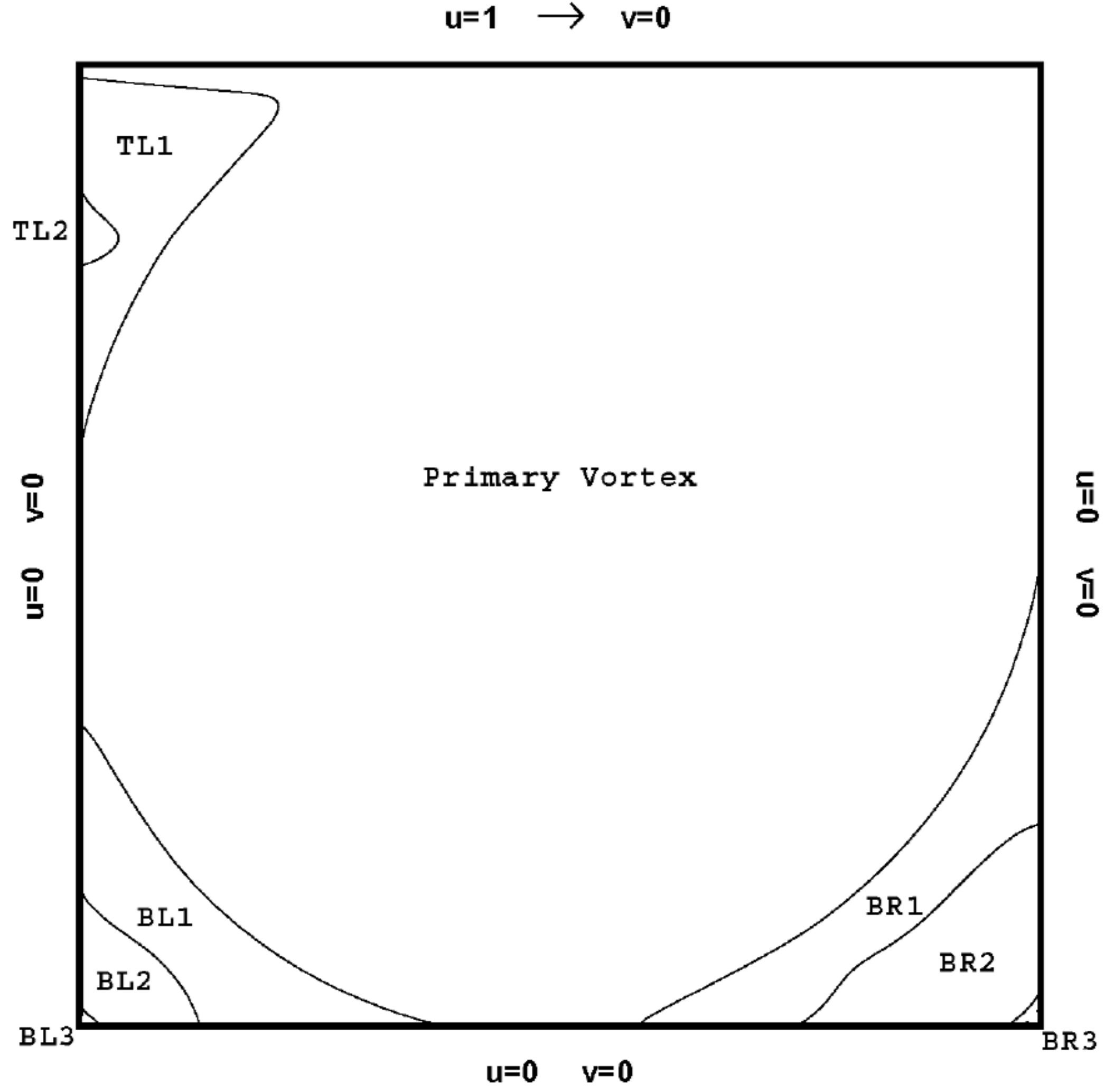}
\par\end{centering}

\caption{Schematic view of the driven lid cavity flow. Source: Erturk \cite{Erturk04}.}

\label{fig:vorticelabels}
\end{figure}

The figure \ref{fig:vorticelabels} shows the boundary conditions
and a schematic of the vortices generated in the driven cavity flow.
In this figure, the abbreviations TL, BL and BR refer to top left,
bottom left and bottom right corners of the cavity, respectively,
and the number following these abbreviations to the vortices that
appear in the flow, numbered according to size, Erturk \cite{Erturk04}.

\begin{figure}[H]
\begin{centering}
\subfloat[\label{fig:5k_S}]{\begin{centering}
\includegraphics[width=10cm]{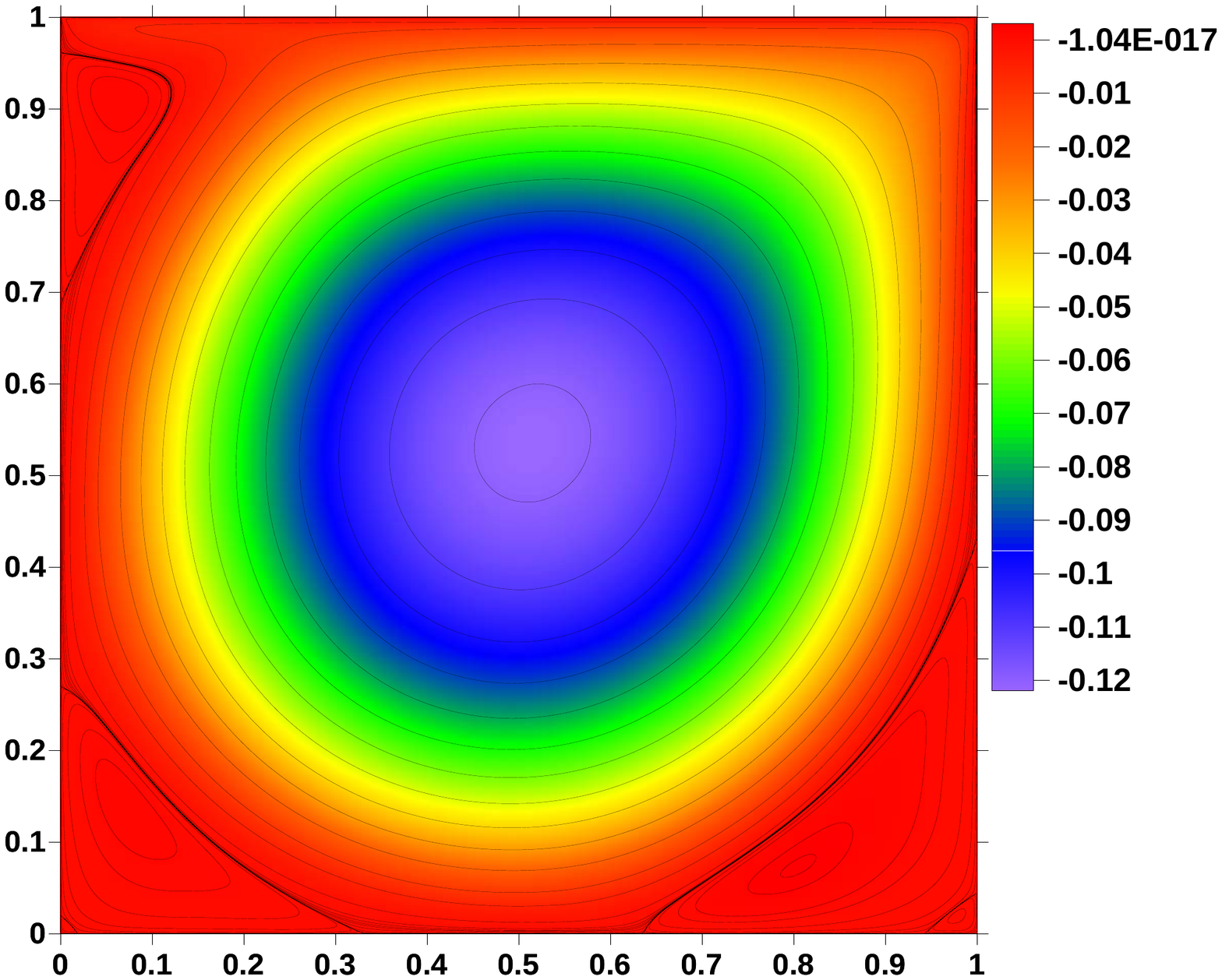}
\par\end{centering}

}
\par\end{centering}

\begin{centering}
\subfloat[\label{fig:5k_S_BL3}]{\begin{centering}
\includegraphics[width=7cm]{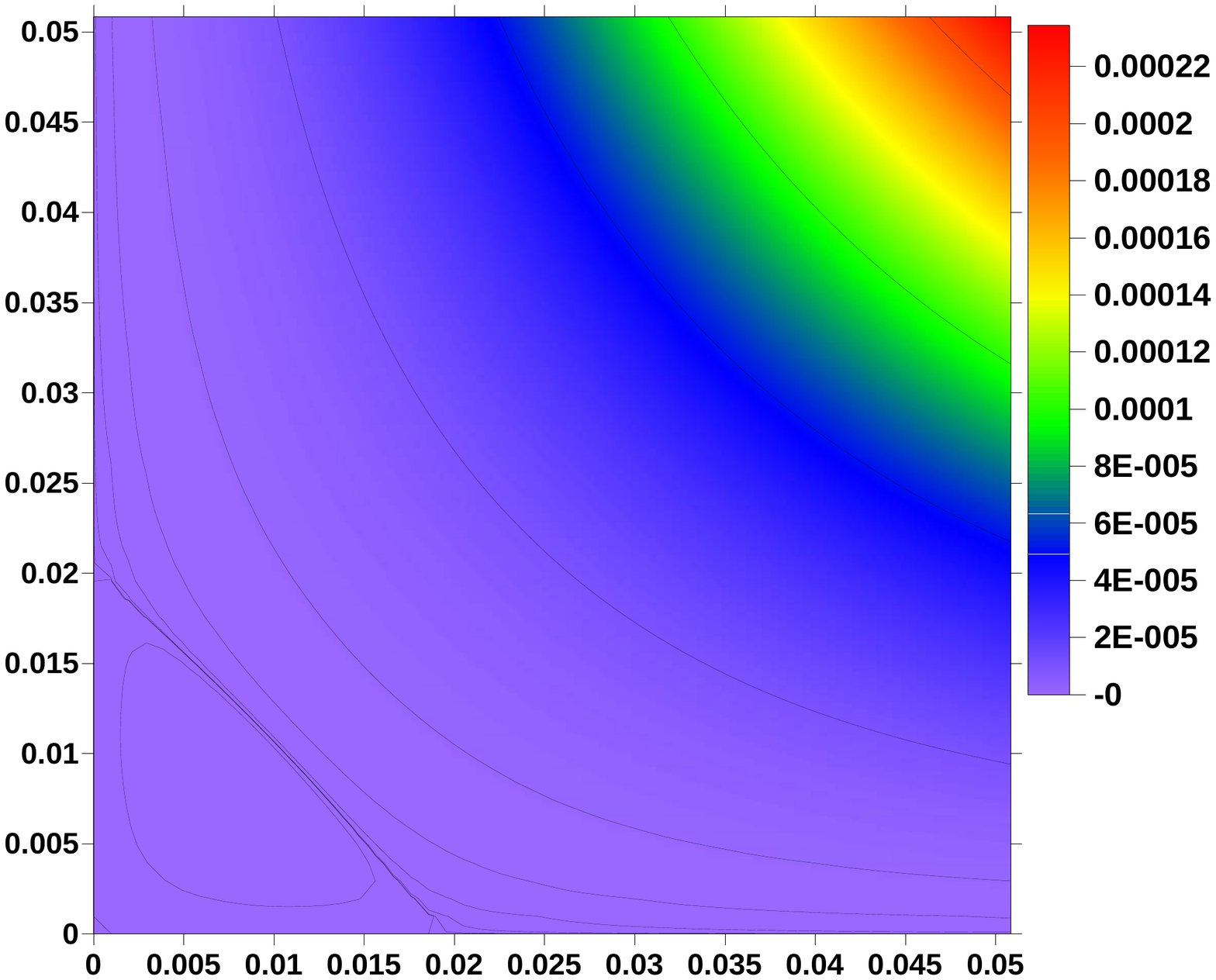}
\par\end{centering}

}\subfloat[\label{fig:5k_S_BR3}]{\begin{centering}
\includegraphics[width=7cm]{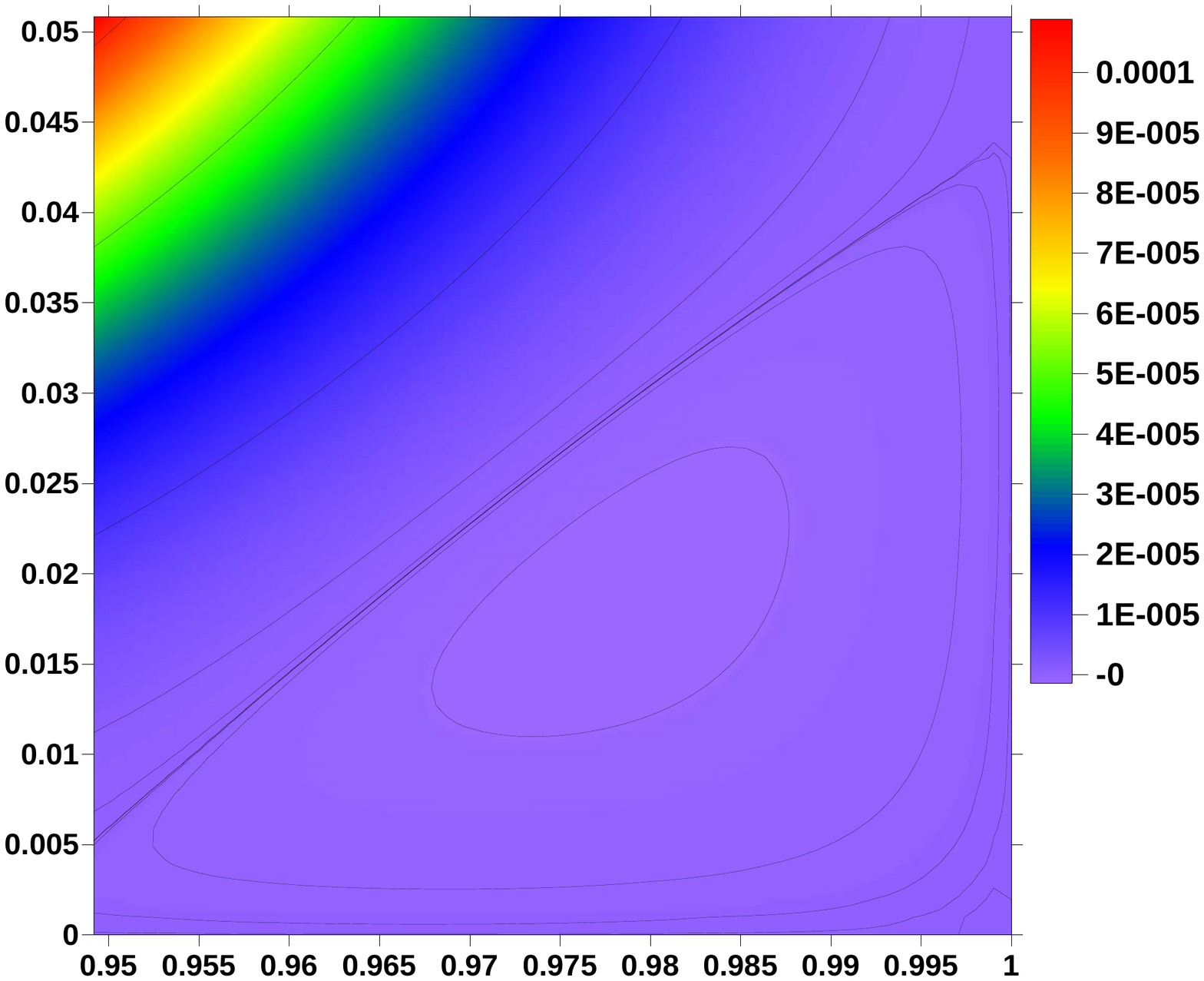}
\par\end{centering}

}
\par\end{centering}

\caption{\protect\subref{fig:5k_S} Stream function contours for $\re=5\,000$ with
$1024\times1024$ points. \protect\subref{fig:5k_S_BL3}  and \protect\subref{fig:5k_S_BR3}
correspond to the Stream function in the bottom left and right corners
of the cavity respectively. The stream function is expressed in $m^{2}s^{-1}$
and the sides of the cavity in $m$.\label{fig:5kS}}

\end{figure}

\begin{figure}[H]
\noindent \begin{centering}
\subfloat[\label{fig:10k_S}]{\begin{centering}
\includegraphics[width=10cm]{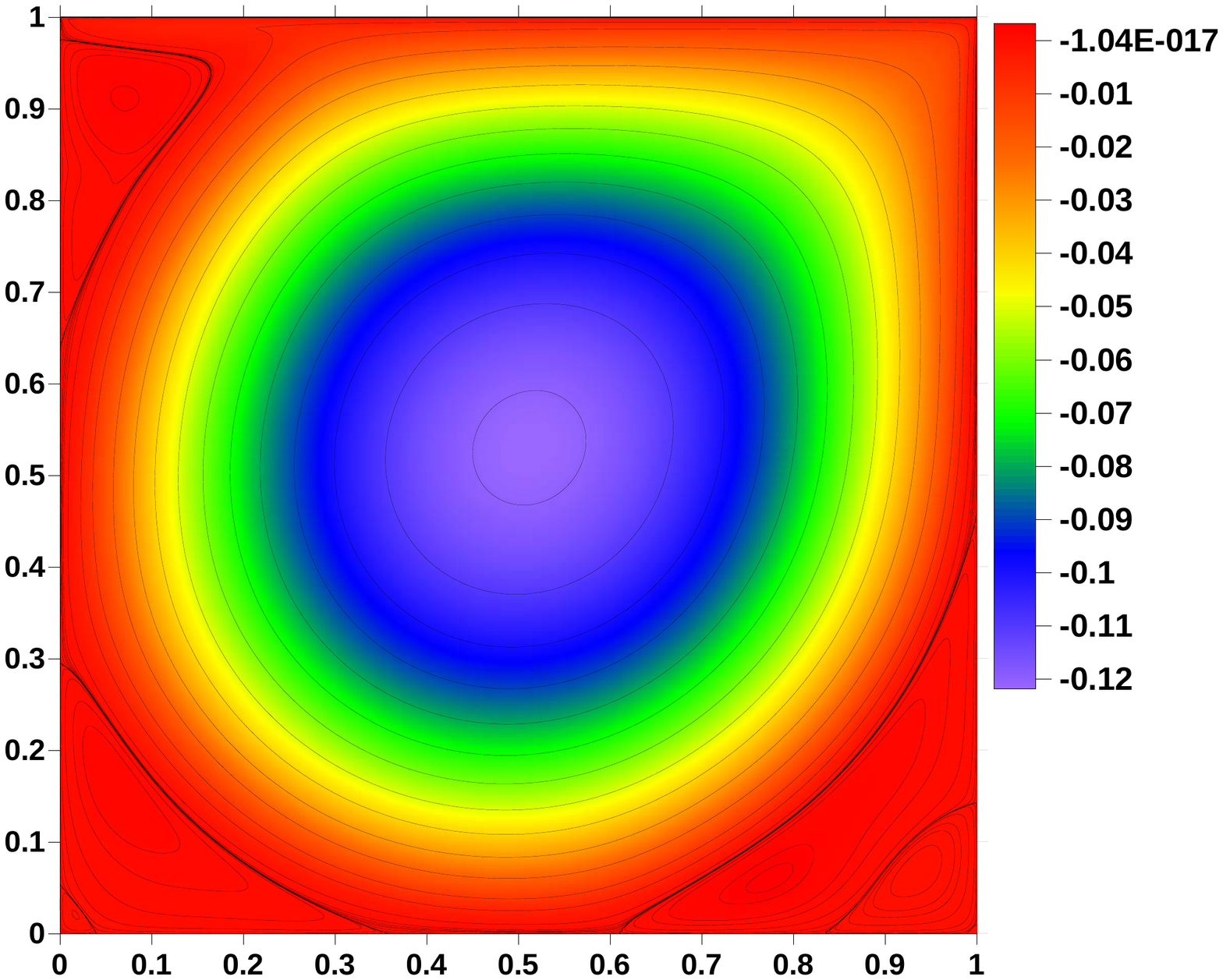}
\par\end{centering}

}
\par\end{centering}

\noindent \begin{centering}
\subfloat[\label{fig:10k_S_BL3}]{\begin{centering}
\includegraphics[width=7cm]{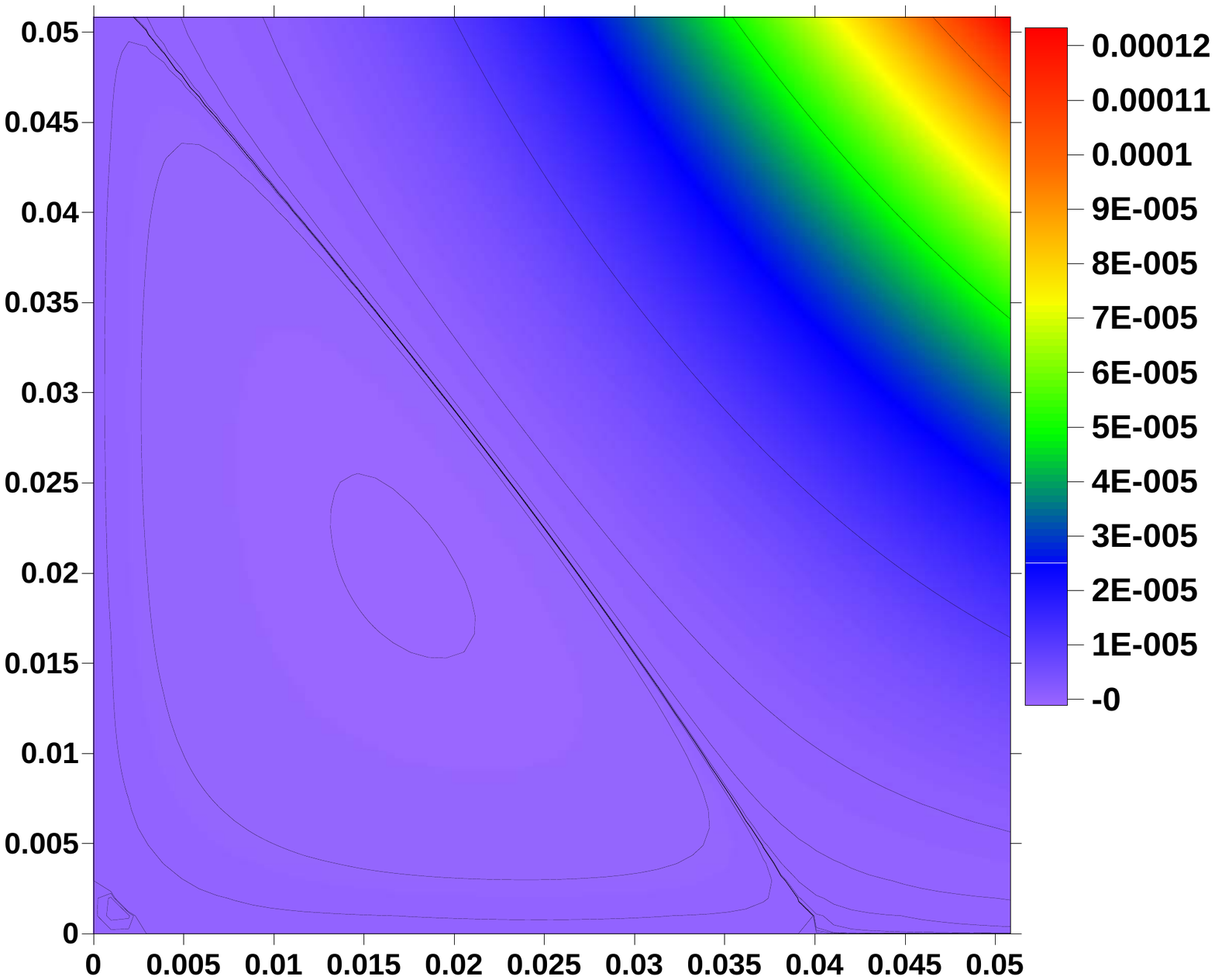}
\par\end{centering}

}\subfloat[\label{fig:10k_S_BR3}]{\begin{centering}
\includegraphics[width=7cm]{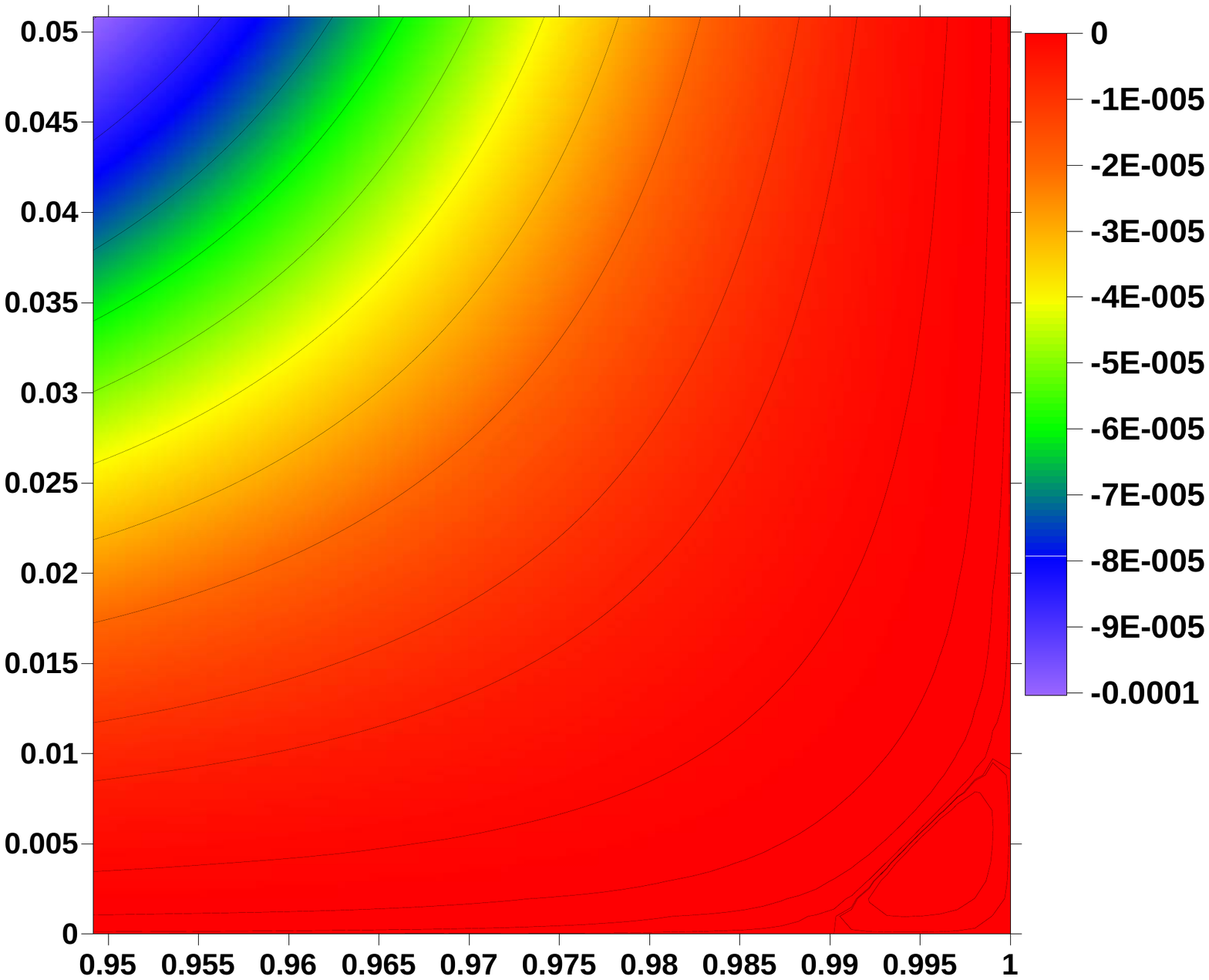}
\par\end{centering}

}
\par\end{centering}

\caption{\protect\subref{fig:10k_S} Stream function contours for $\re=10\,000$ with
$1024\times1024$ points. \protect\subref{fig:10k_S_BL3}  and \protect\subref{fig:10k_S_BR3}
correspond to the Stream function in the bottom left and right corners
of the cavity respectively. The stream function is expressed in $m^{2}s^{-1}$
and the sides of the cavity in $m$.\label{fig:10kS}}

\end{figure}

\begin{figure}[H]
\begin{centering}
\subfloat[\label{fig:15k_S}]{\begin{centering}
\includegraphics[width=10cm]{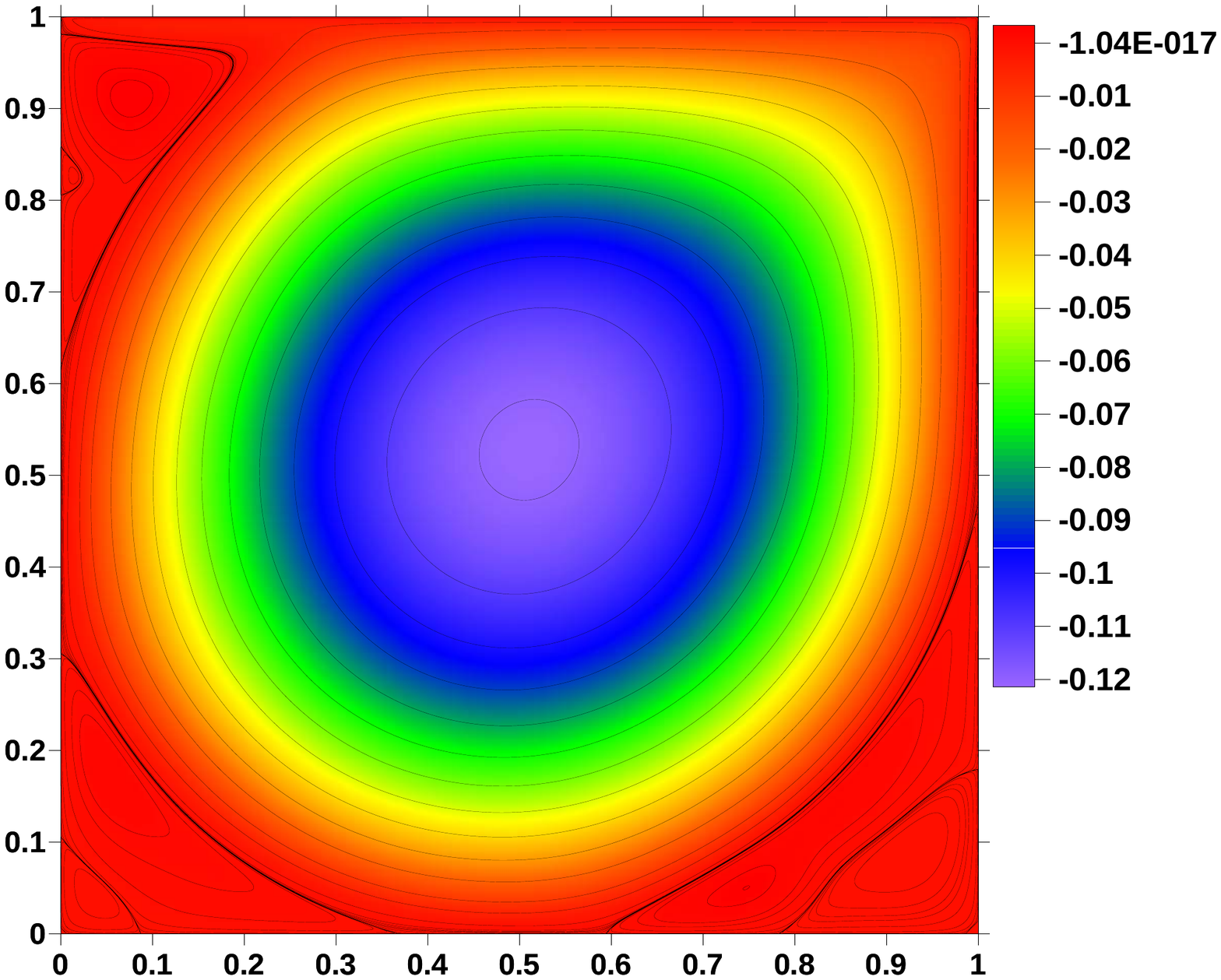}
\par\end{centering}

}
\par\end{centering}

\begin{centering}
\subfloat[\label{fig:15k_S_BL3}]{\begin{centering}
\includegraphics[width=7cm]{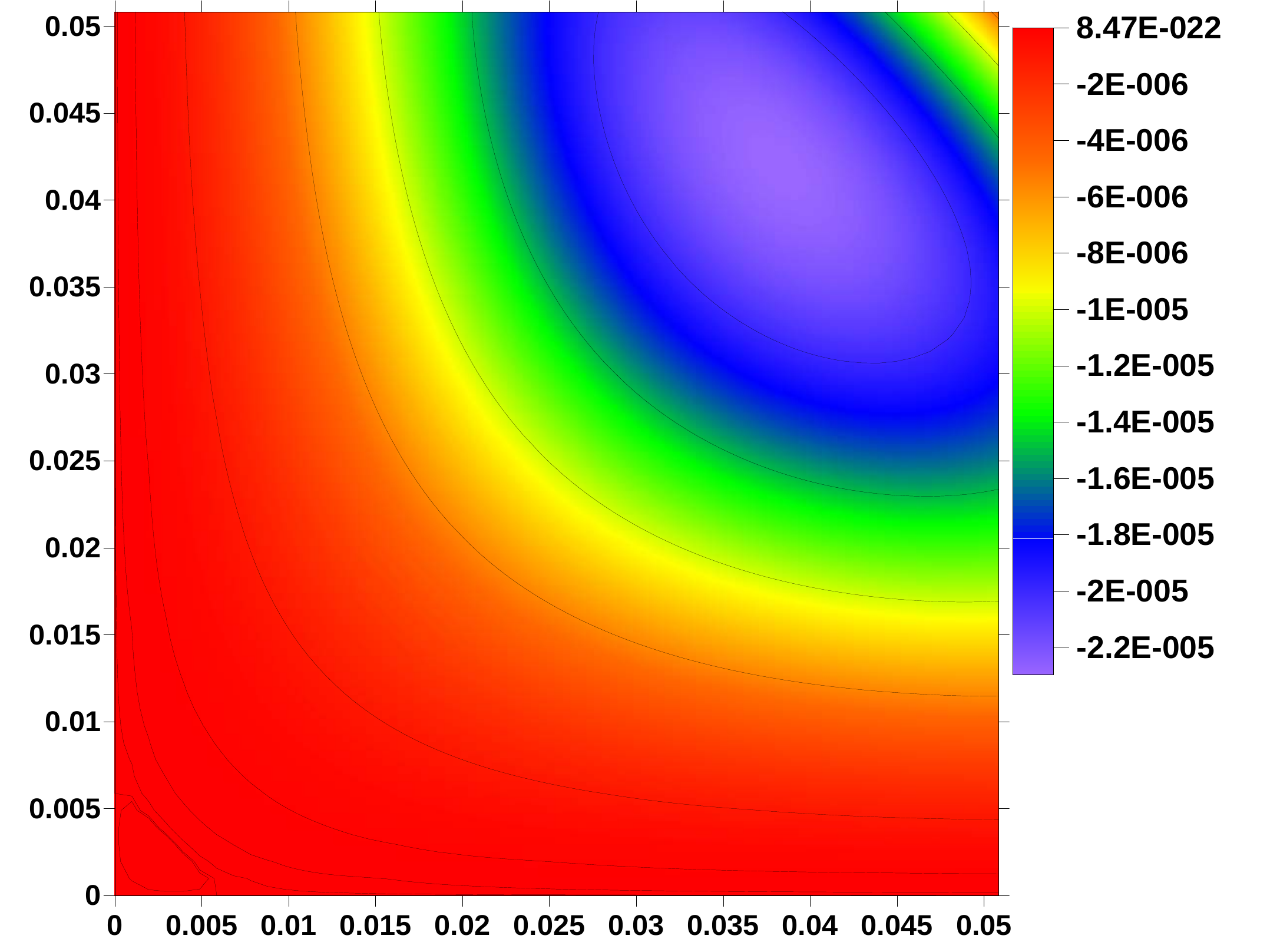}
\par\end{centering}

}\subfloat[\label{fig:15k_S_BR3}]{\begin{centering}
\includegraphics[width=7cm]{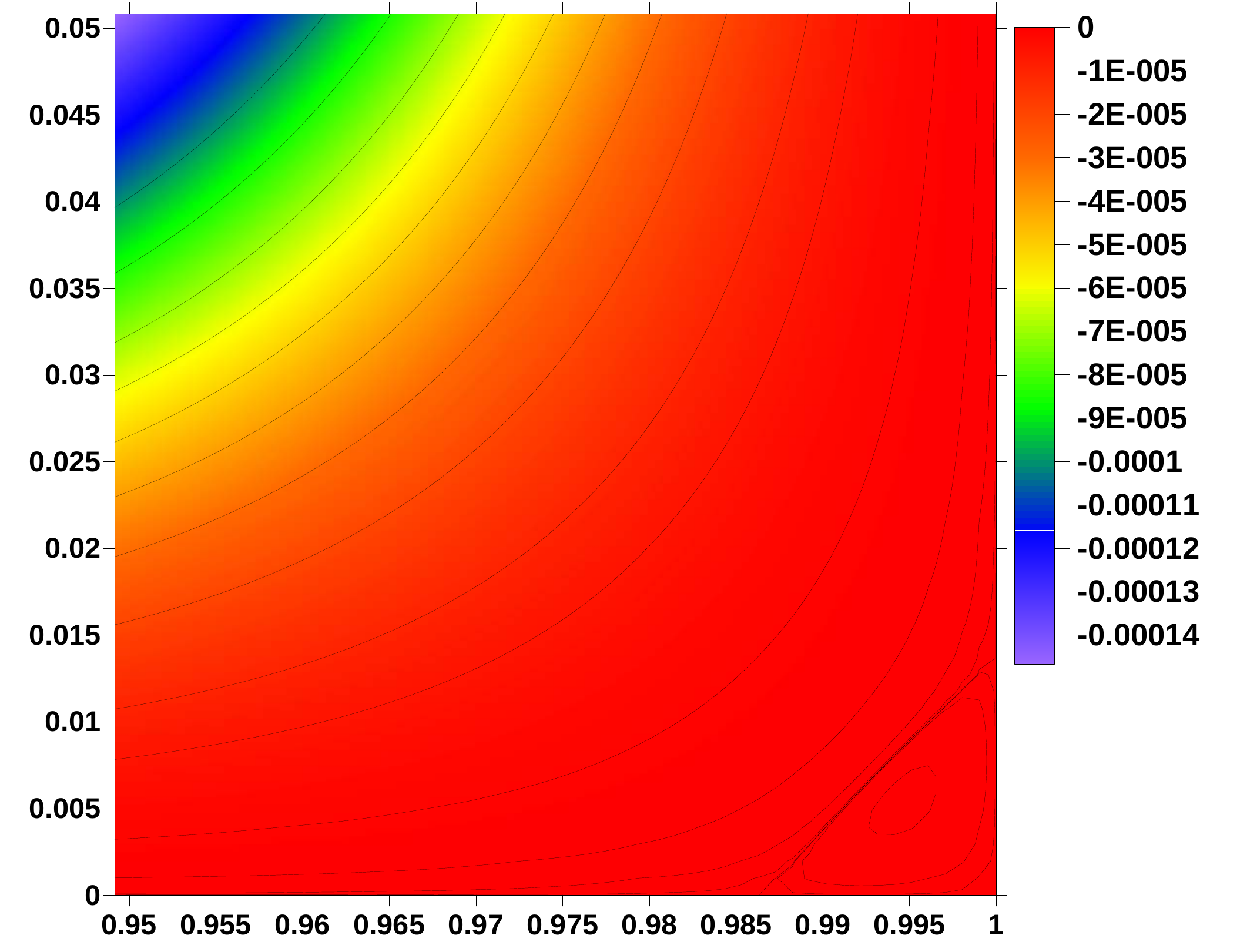}
\par\end{centering}

}
\par\end{centering}

\caption{\protect\subref{fig:15k_S} Stream function contours for $\re=15\,000$ with
$1024\times1024$ points. \protect\subref{fig:15k_S_BL3}  and \protect\subref{fig:15k_S_BR3}
correspond to the Stream function in the bottom left and right corners
of the cavity respectively. The stream function is expressed in $m^{2}s^{-1}$
and the sides of the cavity in $m$.\label{fig:15kS}}

\end{figure}

\begin{figure}[H]
\noindent \begin{centering}
\subfloat[\label{fig:20k_S}]{\begin{centering}
\includegraphics[width=10cm]{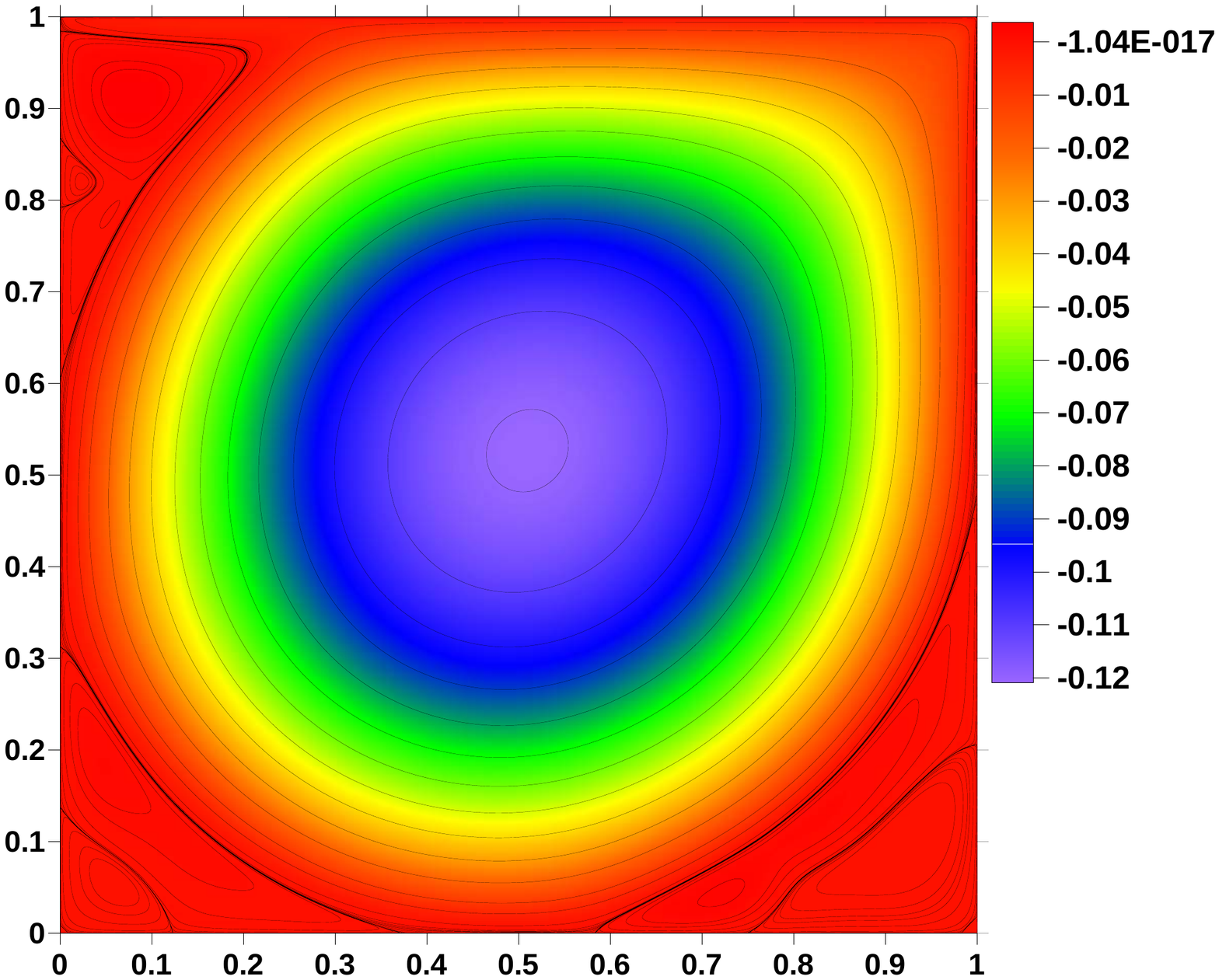}
\par\end{centering}

}
\par\end{centering}

\noindent \begin{centering}
\subfloat[\label{fig:20k_S_BL3}]{\begin{centering}
\includegraphics[width=7cm]{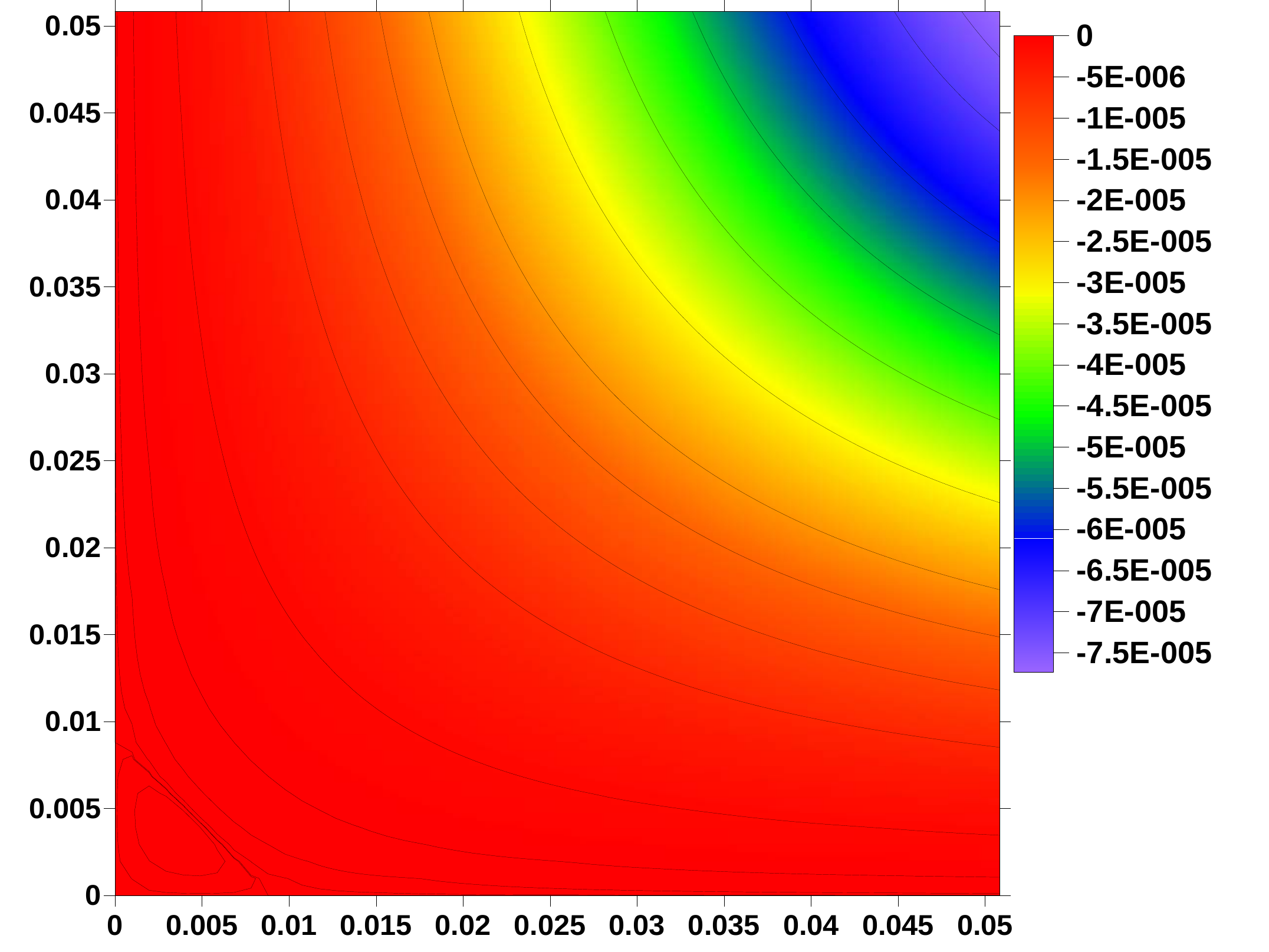}
\par\end{centering}

}\subfloat[\label{fig:20k_S_BR3}]{\begin{centering}
\includegraphics[width=7cm]{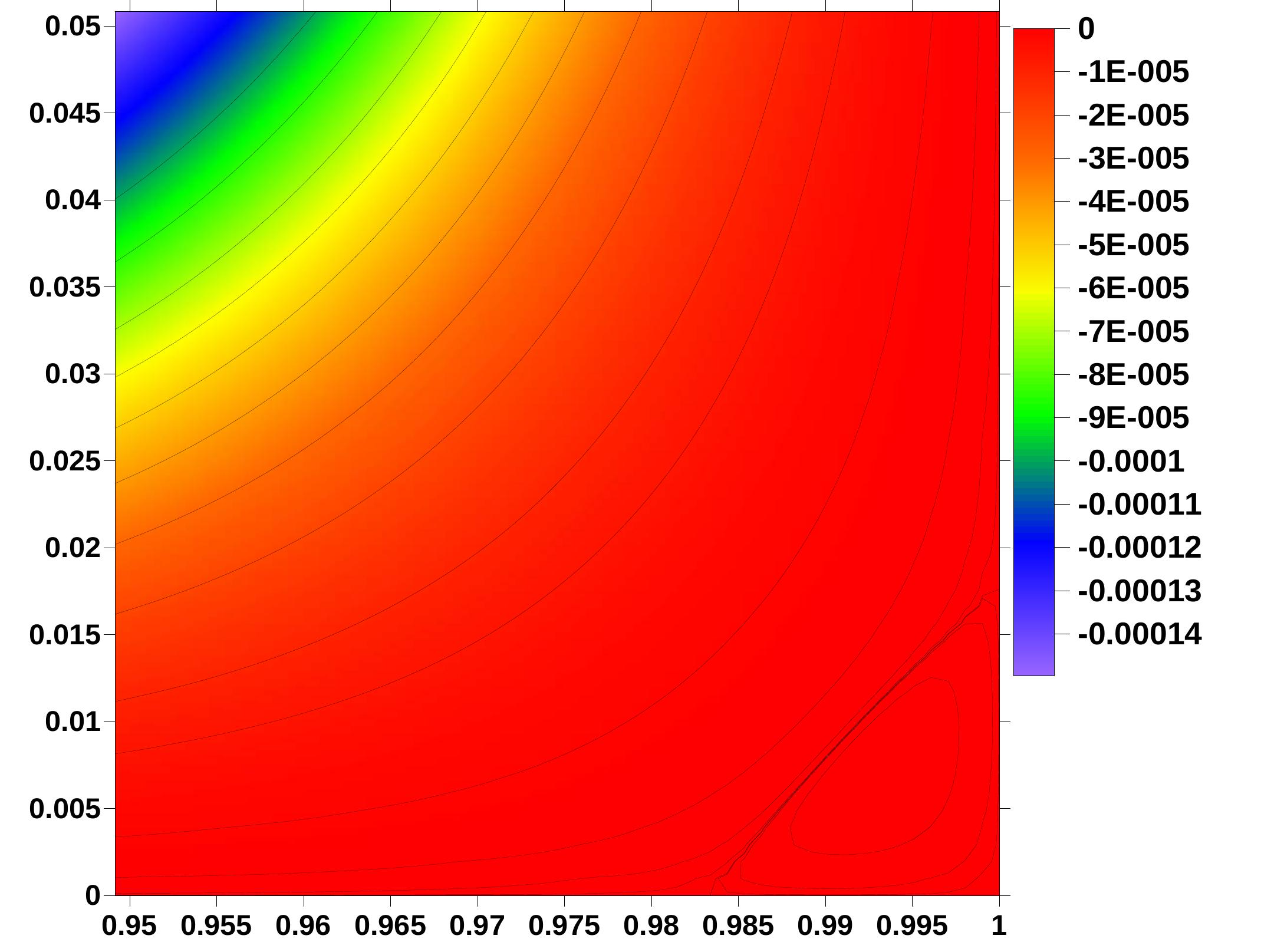}
\par\end{centering}

}
\par\end{centering}

\caption{\protect\subref{fig:20k_S} Stream function contours for $\re=20\,000$ with
$1024\times1024$ points. \protect\subref{fig:20k_S_BL3}  and \protect\subref{fig:20k_S_BR3}
correspond to the Stream function in the bottom left and right corners
of the cavity respectively. The stream function is expressed in $m^{2}s^{-1}$
and the sides of the cavity in $m$.\label{fig:20kS}}

\end{figure}

\begin{figure}[H]
\noindent \begin{centering}
\subfloat[\label{fig:25k_S}]{\begin{centering}
\includegraphics[width=10cm]{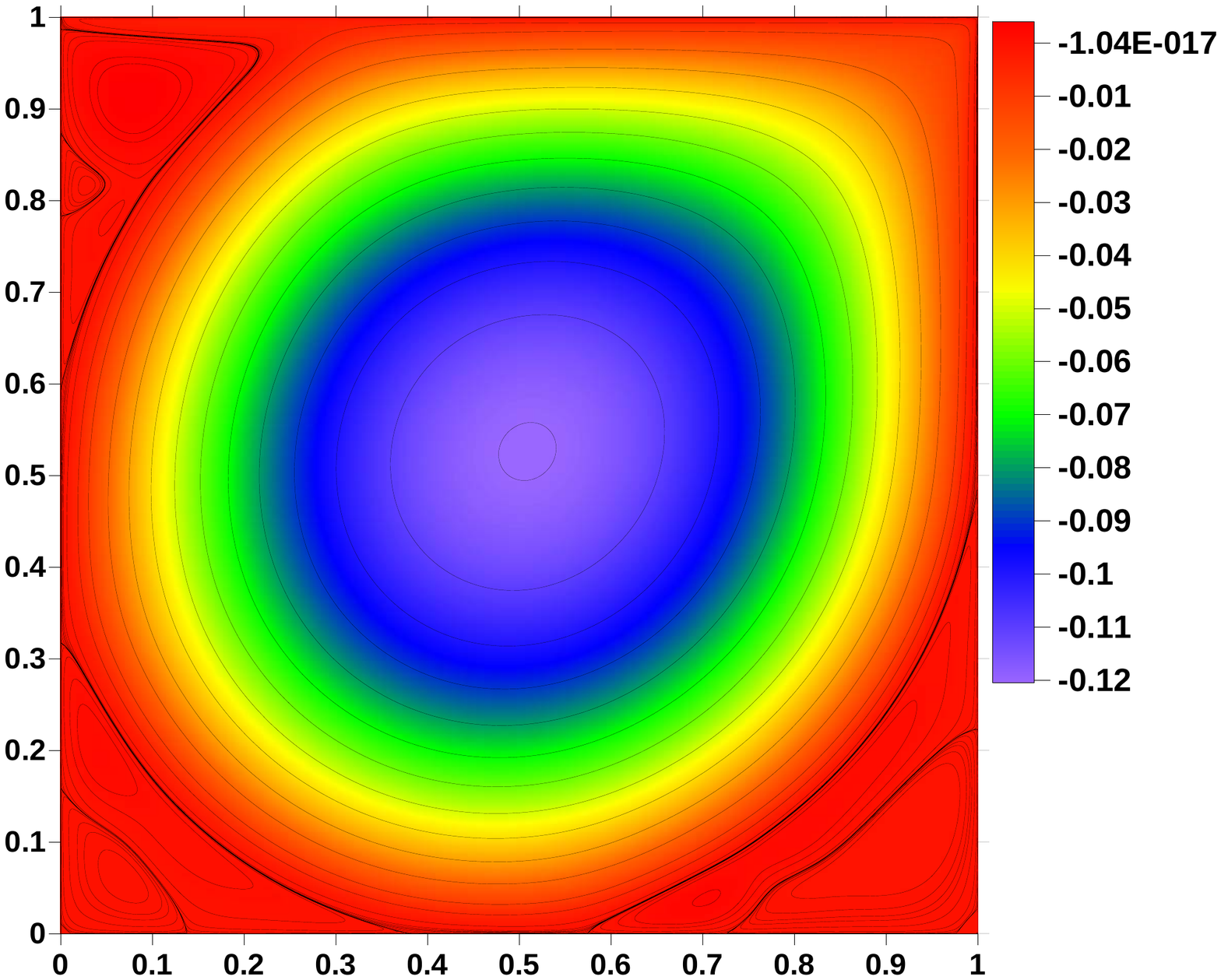}
\par\end{centering}

}
\par\end{centering}

\noindent \begin{centering}
\subfloat[\label{fig:25k_S_BL3}]{\begin{centering}
\includegraphics[width=7cm]{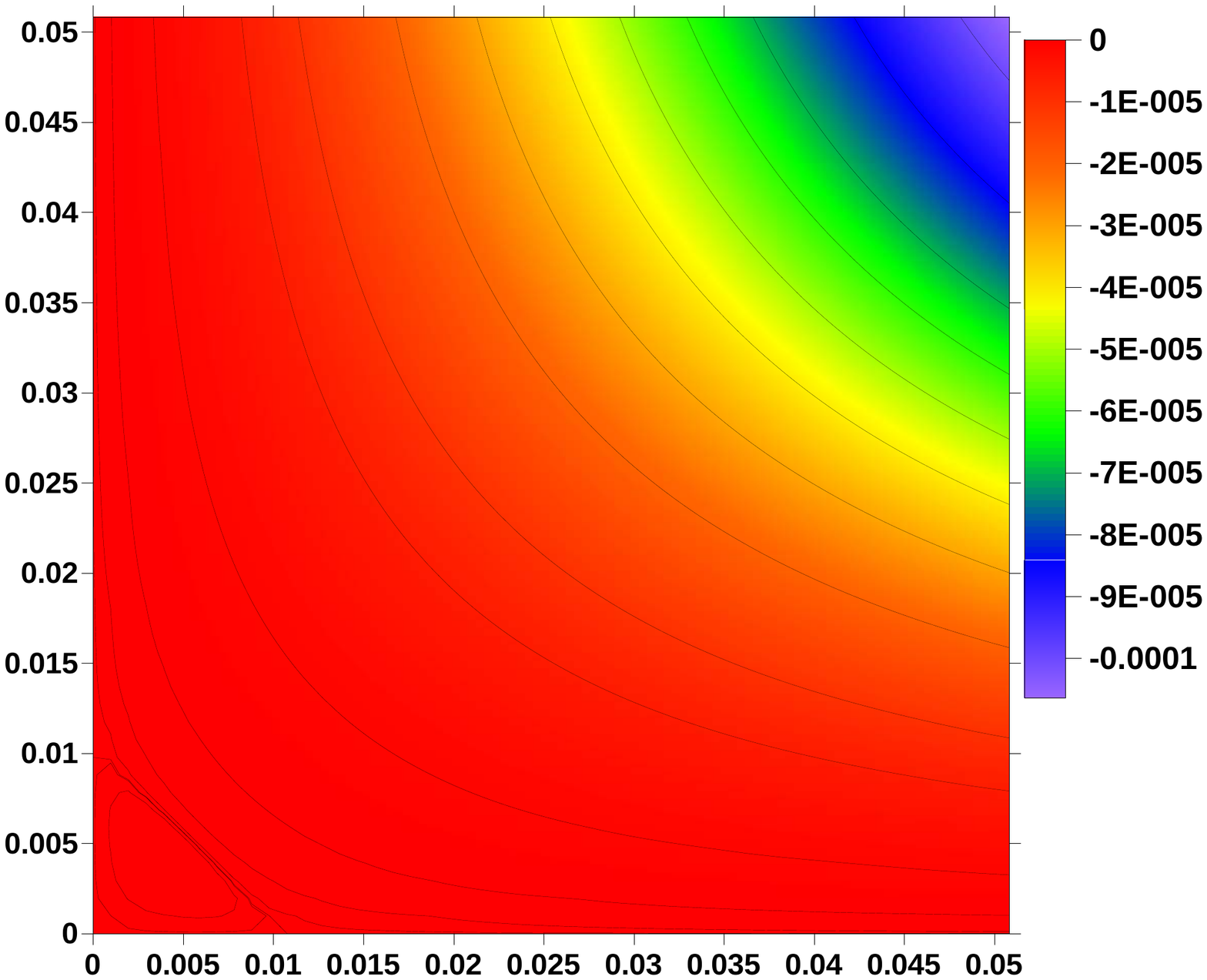}
\par\end{centering}

}\subfloat[\label{fig:25k_S_BR3}]{\begin{centering}
\includegraphics[width=7cm]{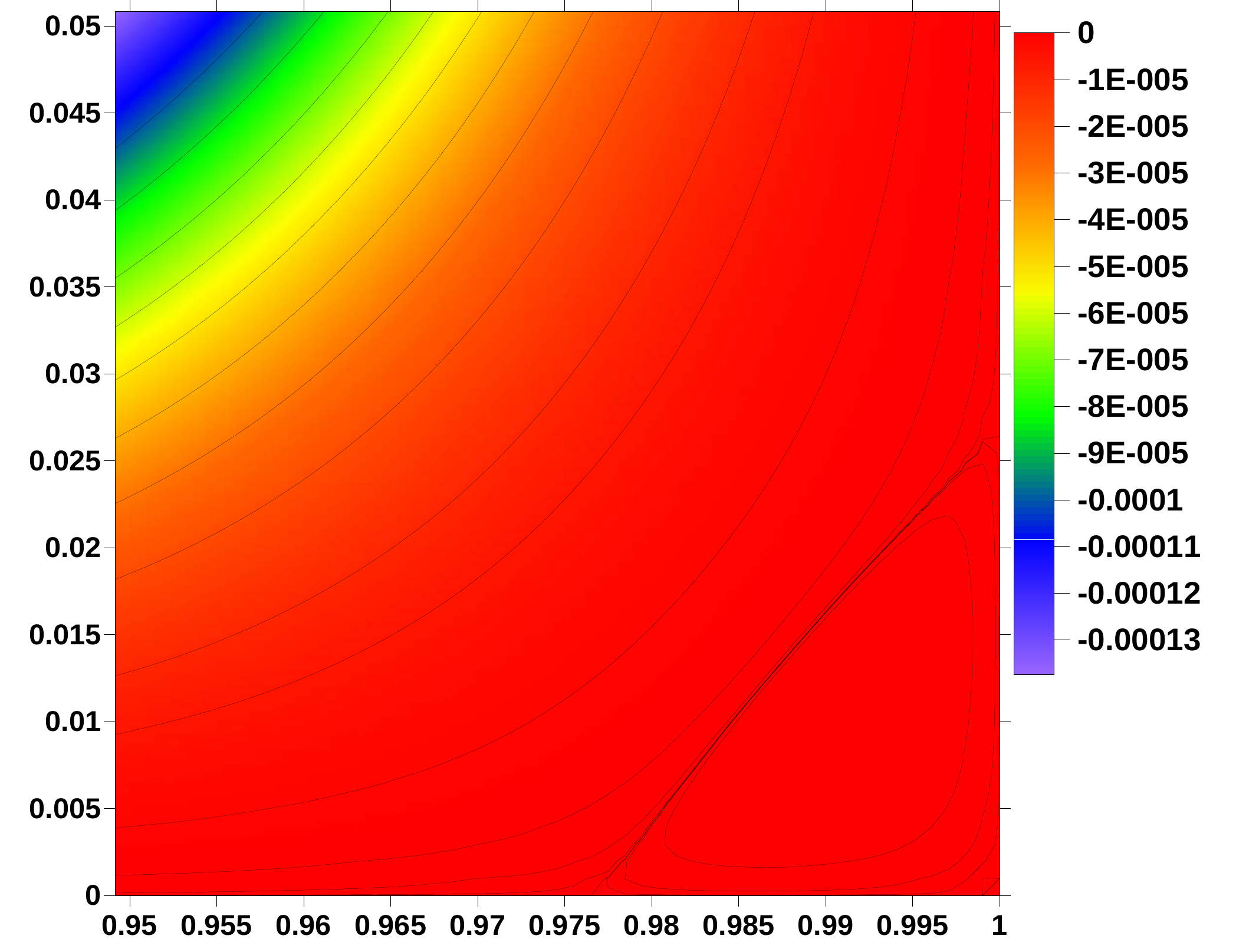}
\par\end{centering}

}
\par\end{centering}

\caption{\protect\subref{fig:25k_S} Stream function contours for $\re=25\,000$ with
$1024\times1024$ points. \protect\subref{fig:25k_S_BL3}  and \protect\subref{fig:25k_S_BR3}
correspond to the Stream function in the bottom left and right corners
of the cavity respectively. The stream function is expressed in $m^{2}s^{-1}$
and the sides of the cavity in $m$.\label{fig:25kS}}

\end{figure}

\begin{figure}[H]
\noindent \begin{centering}
\subfloat[$\re=5\,000$\label{fig:5kV}]{\begin{centering}
\includegraphics[width=8cm]{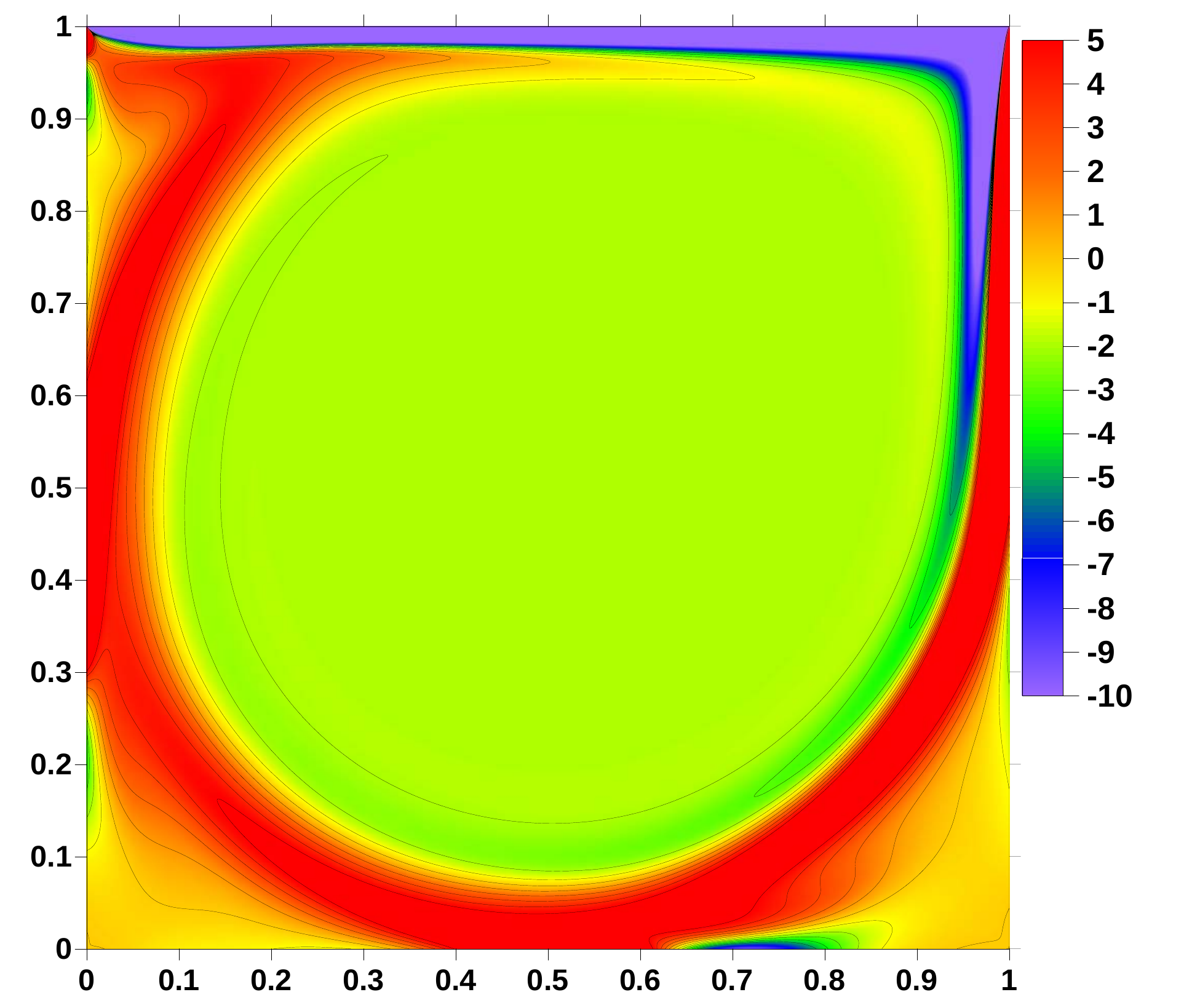}
\par\end{centering}

}\subfloat[$\re=10\,000$\label{fig:10kV}]{\begin{centering}
\includegraphics[width=8cm]{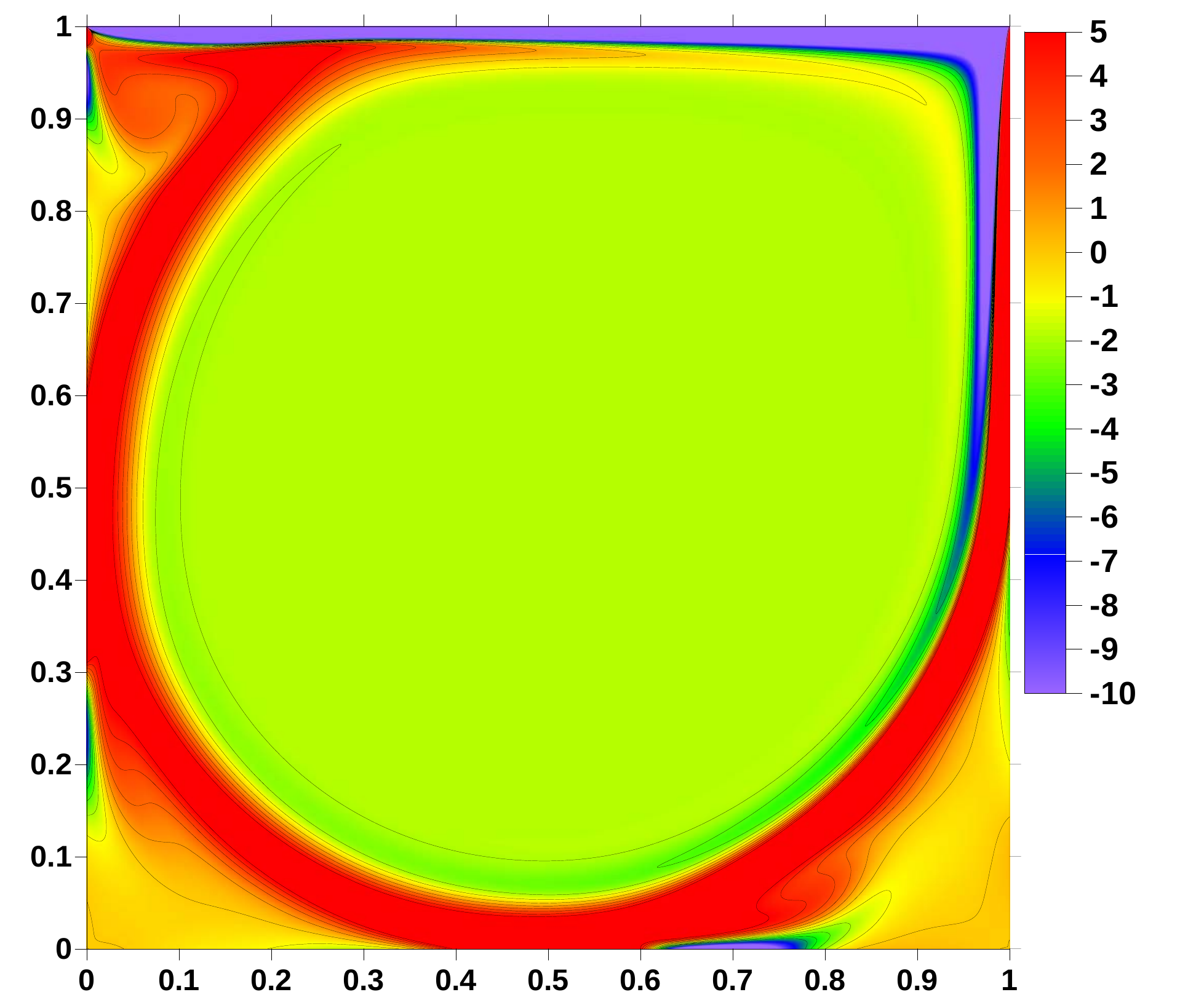}
\par\end{centering}

}
\par\end{centering}

\noindent \begin{centering}
\subfloat[$\re=15\,000$\label{fig:15kV}]{\begin{centering}
\includegraphics[width=8cm]{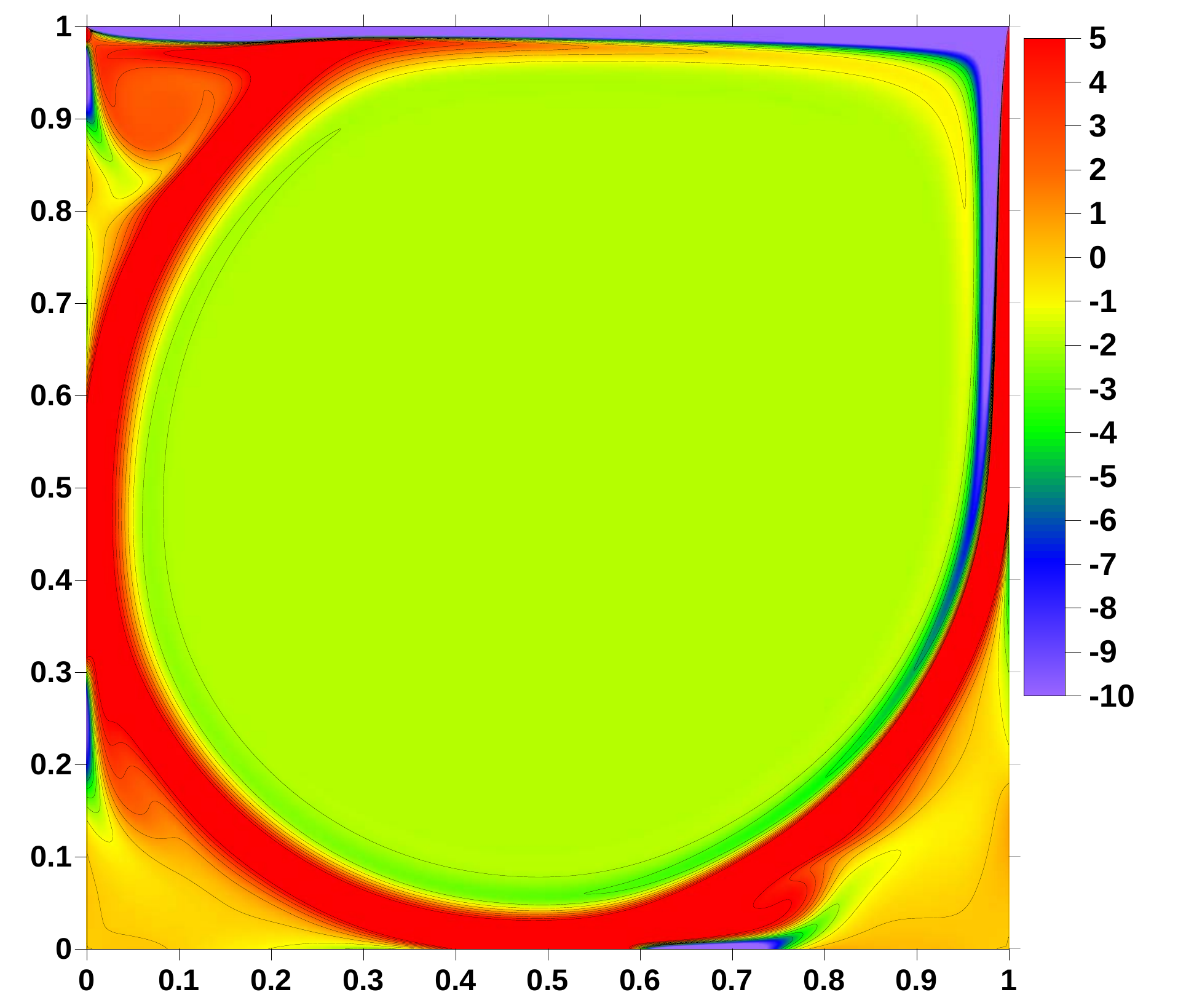}
\par\end{centering}

}\subfloat[$\re=20\,000$\label{fig:20kV}]{\begin{centering}
\includegraphics[width=8cm]{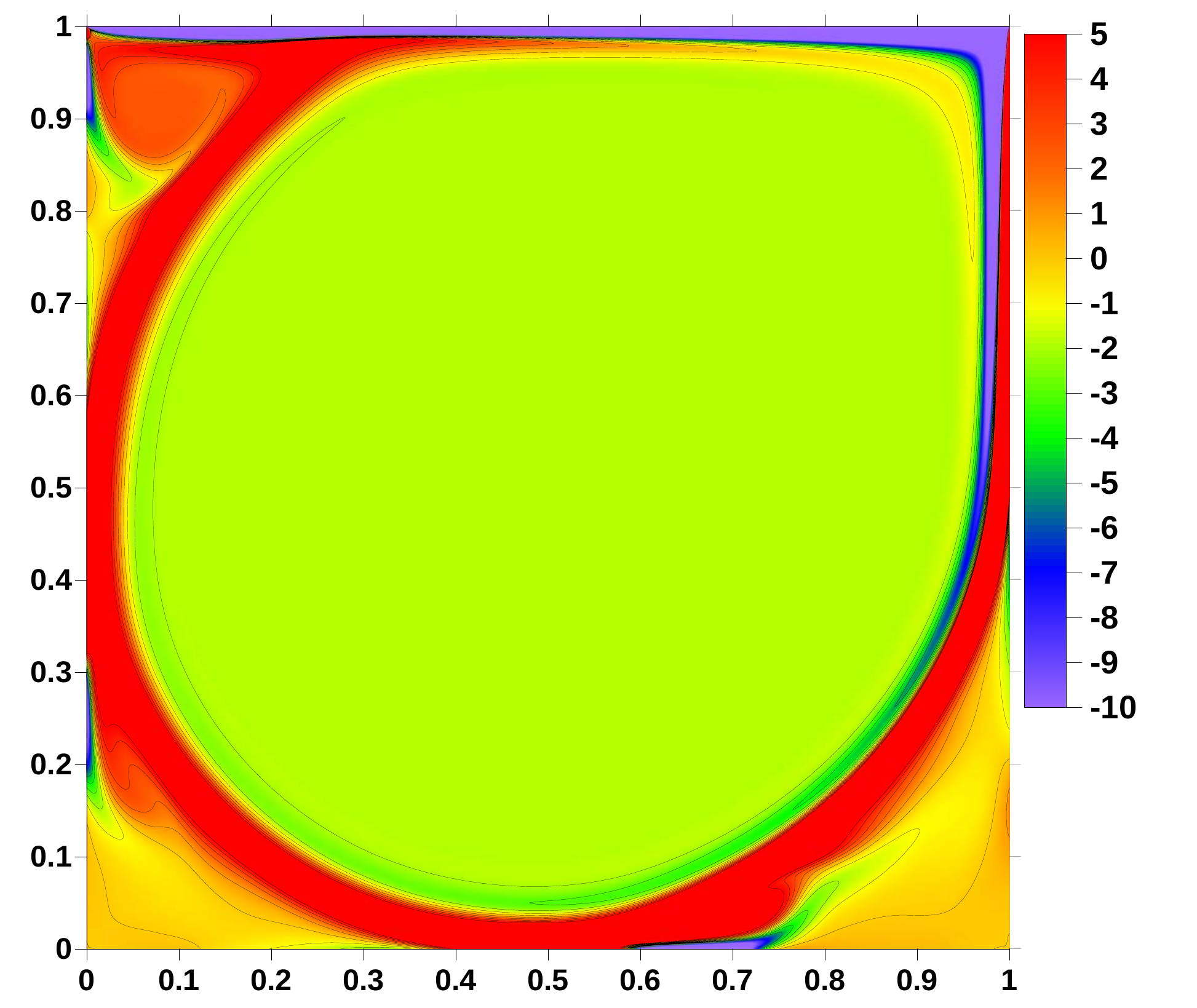}
\par\end{centering}

}
\par\end{centering}

\noindent \begin{centering}
\subfloat[$\re=25\,000$\label{fig:25kV}]{\begin{centering}
\includegraphics[width=8cm]{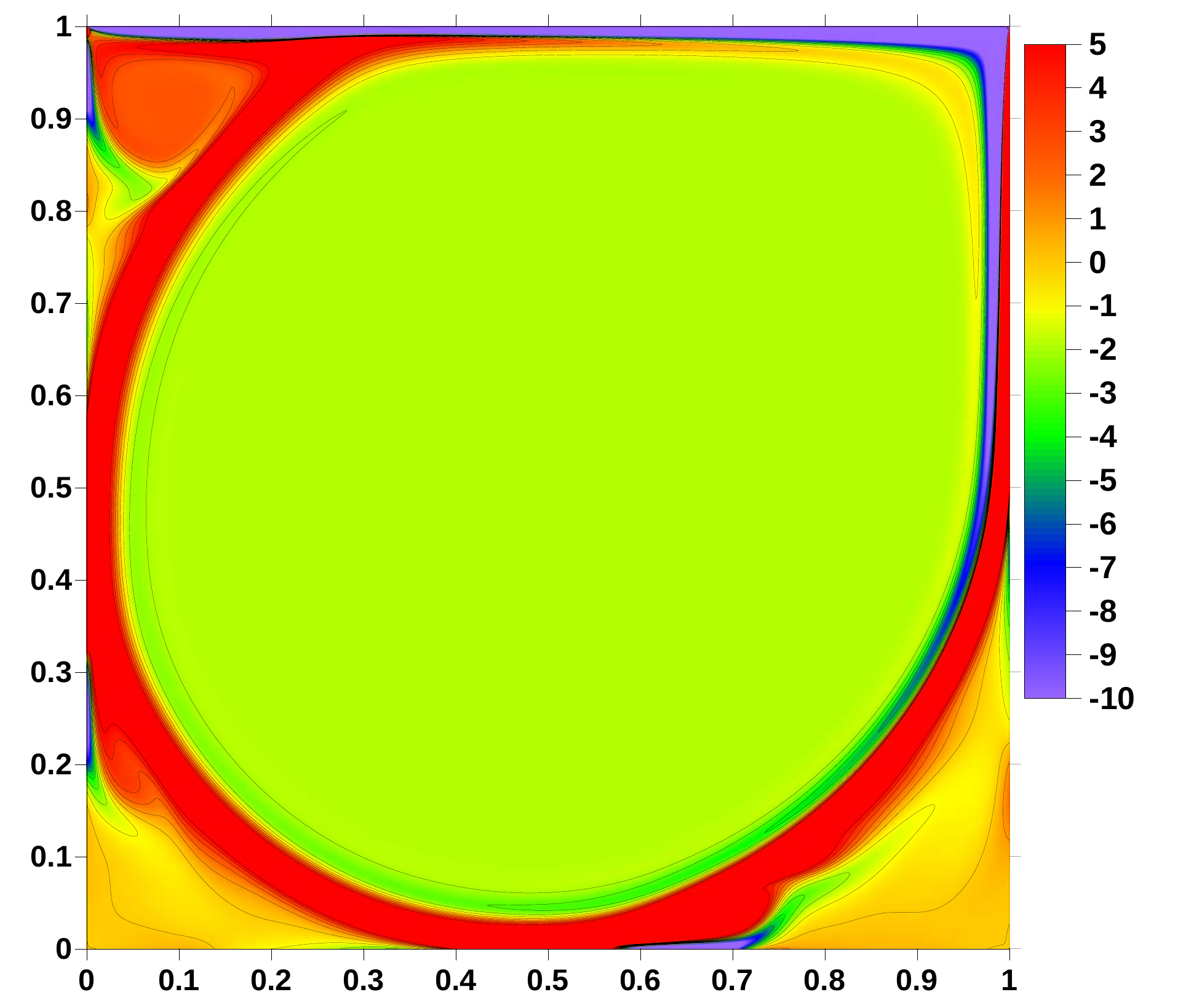}
\par\end{centering}

}\subfloat[$\re=30\,000$\label{fig:30kV}]{\begin{centering}
\includegraphics[width=8cm]{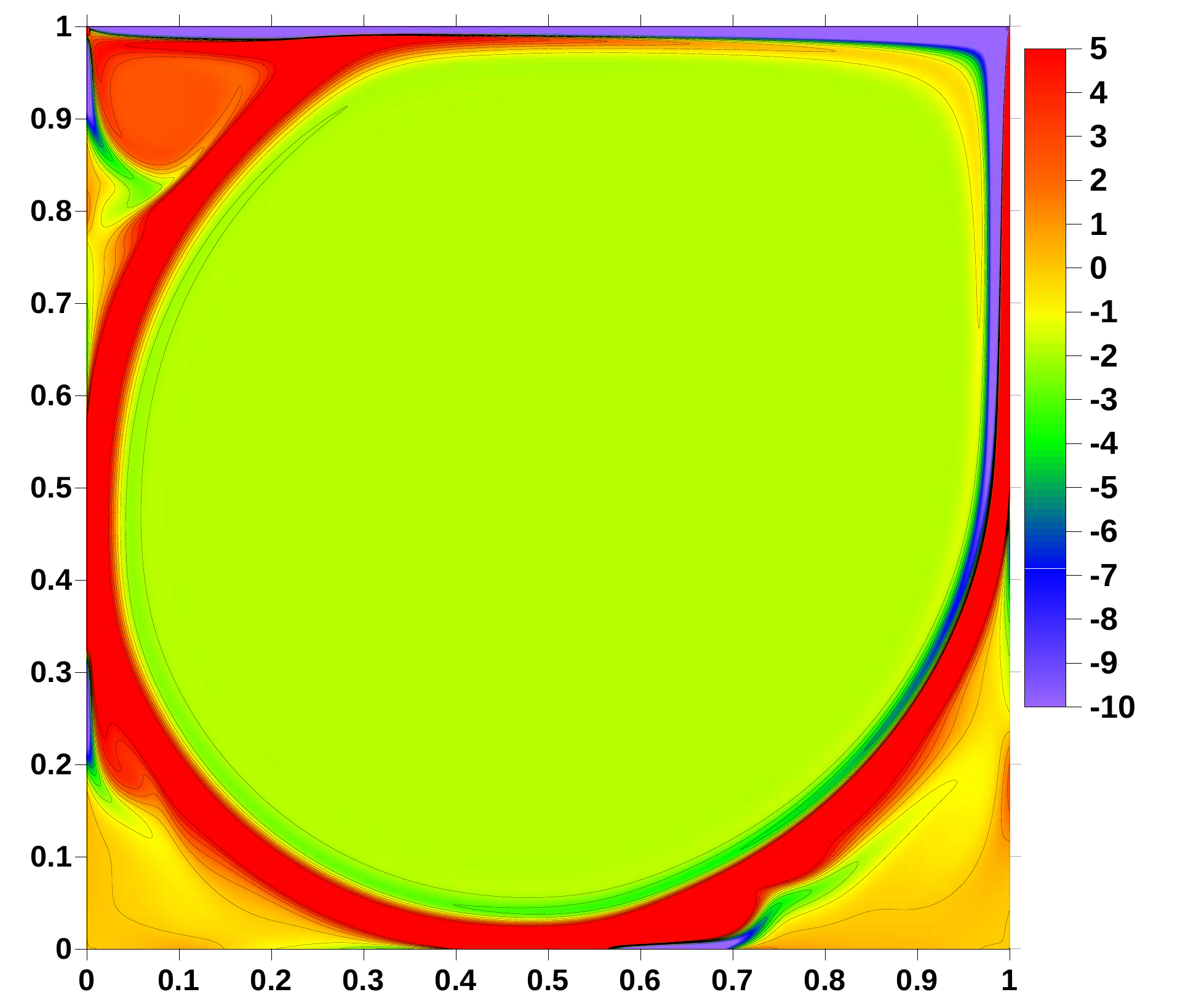}
\par\end{centering}

}
\par\end{centering}

\caption{Vorticity contours at various Reynolds numbers. The vorticity is expressed
in $s^{-1}$ and the sides of the cavity in $m$.\label{fig:figures_Vort_All}}

\end{figure}

Figures \ref{fig:5kS} to \ref{fig:25k_S} show the stream function
contours and Figure \ref{fig:figures_Vort_All} the vorticity contours
of the cavity flow up to $\re=30\,000$. In the figure \ref{fig:5kS}
are represented the stream function contours for $\re=5\,000$, according
to figure \ref{fig:vorticelabels}, all the vortices appear except
the BL3 and TL2. This result is in agreement with the result obtained
by Erturk \cite{Erturk08}. For $\re=10\,000$, figure \ref{fig:10kS},
all the vortices appear except the TL2. For $\re\geq15\,000$, all
the vortices can be seen. These contour figures show that, the fine
grid mesh provides very smooth solutions at high Reynolds numbers.

\begin{figure}[H]
\noindent \begin{centering}
\subfloat[\label{fig:30k_S}]{\begin{centering}
\includegraphics[width=10cm]{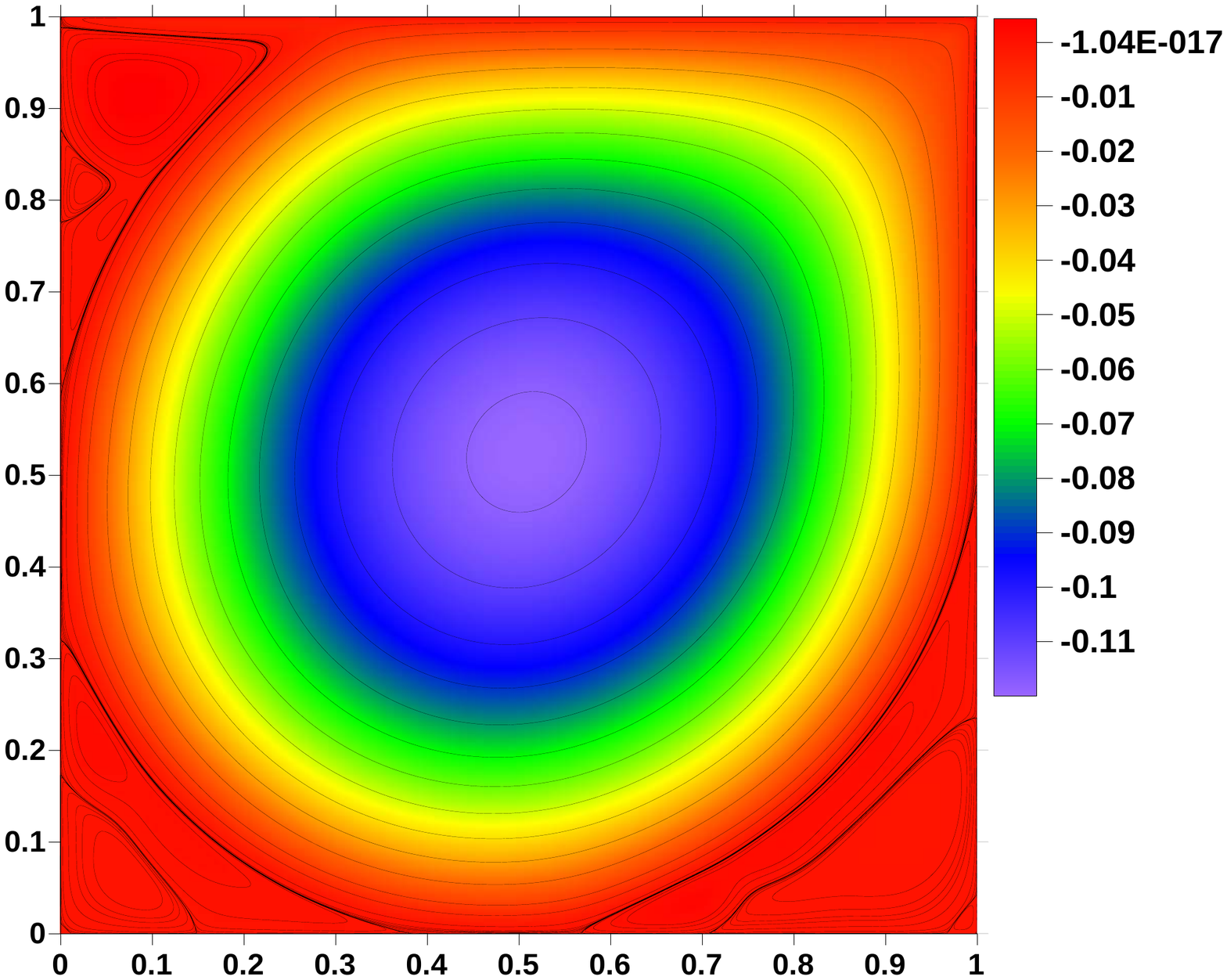}
\par\end{centering}

}
\par\end{centering}

\noindent \begin{centering}
\subfloat[\label{fig:30k_S_BL3}]{\begin{centering}
\includegraphics[width=7cm]{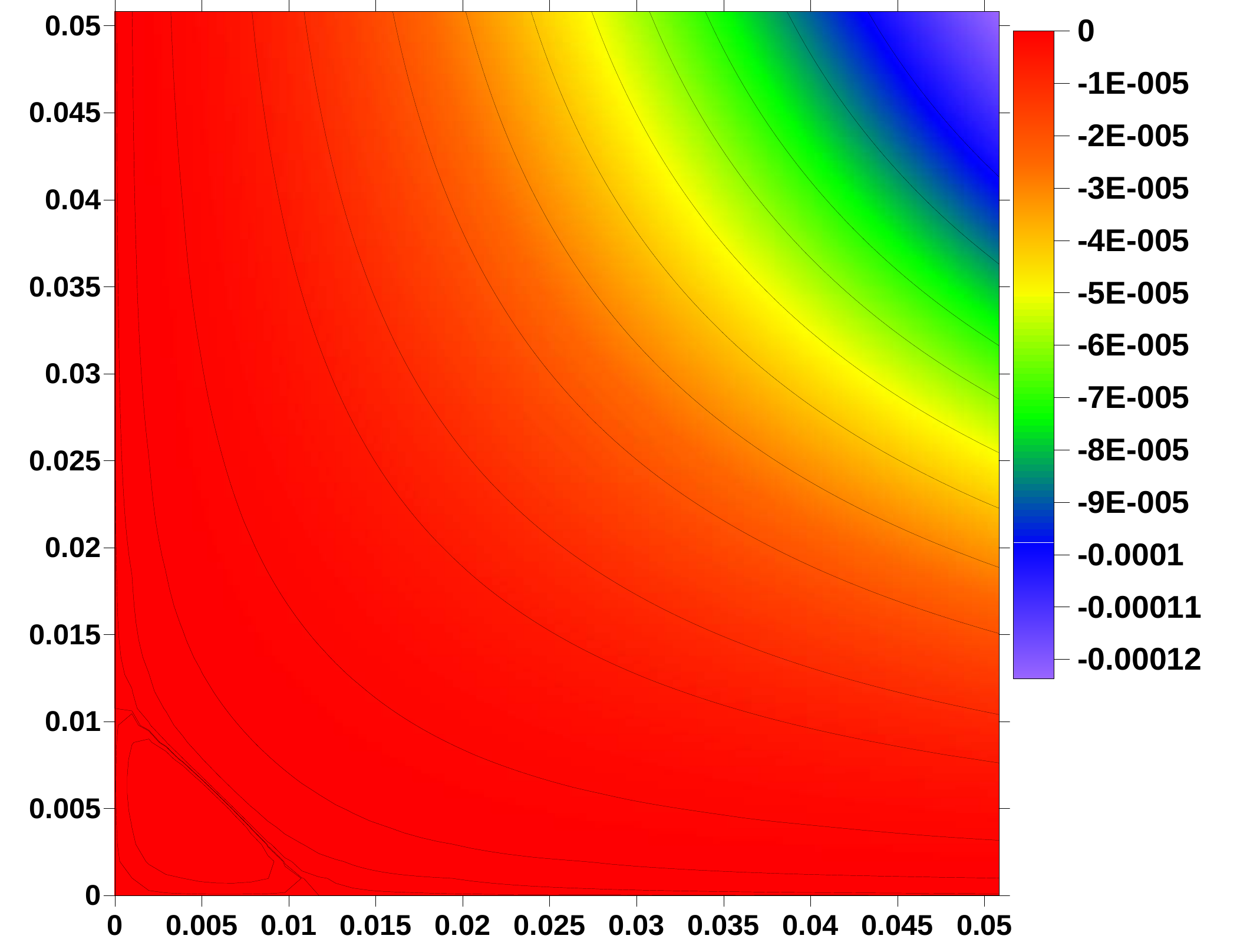}
\par\end{centering}

}\subfloat[\label{fig:30k_S_BR3}]{\begin{centering}
\includegraphics[width=7cm]{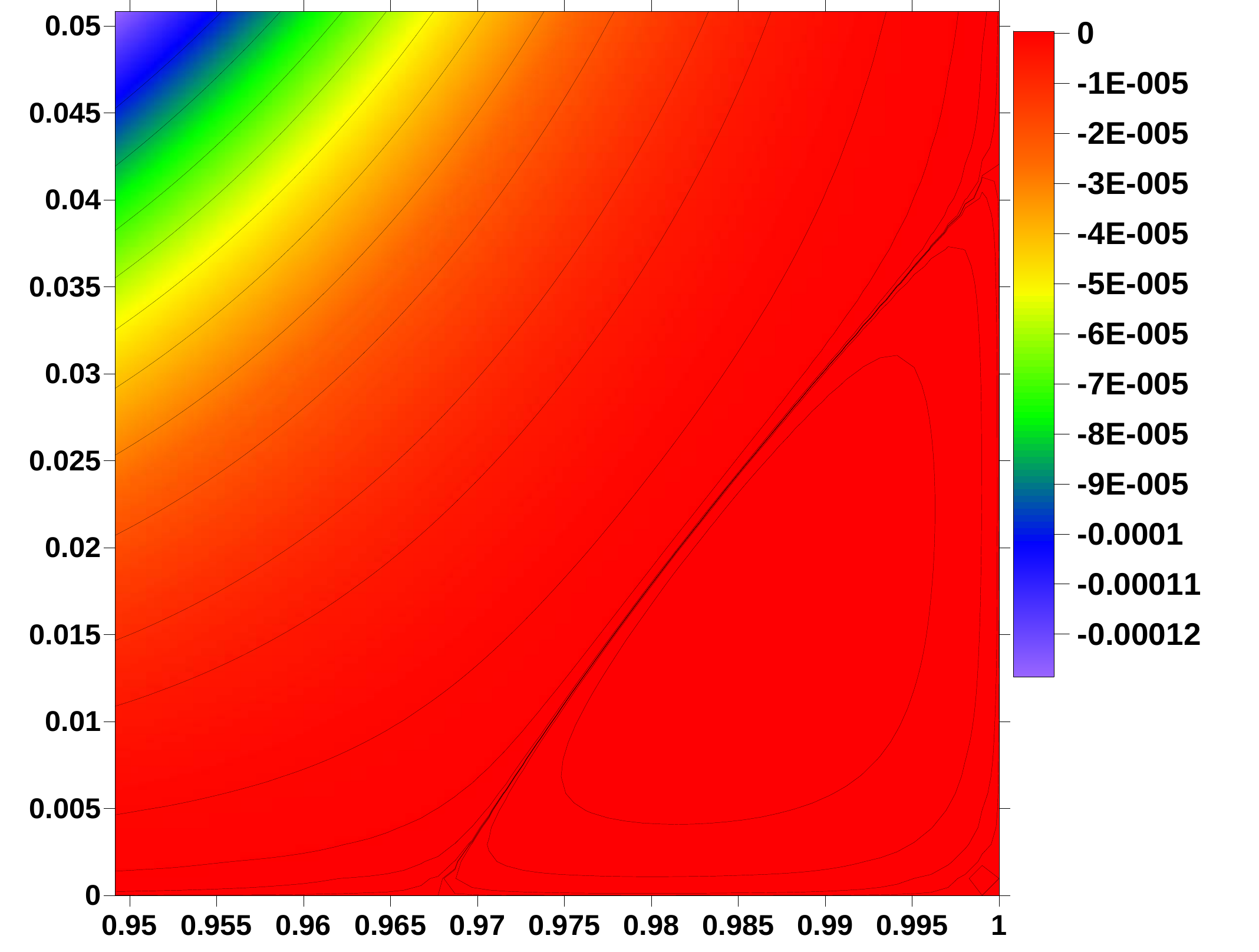}
\par\end{centering}

}
\par\end{centering}

\caption{\protect\subref{fig:30k_S} Stream function contours for $\re=30\,000$ with
$1024\times1024$ points. \protect\subref{fig:30k_S_BL3}  and \protect\subref{fig:30k_S_BR3}
correspond to the Stream function in the bottom left and right corners
of the cavity respectively. The stream function is expressed in $m^{2}s^{-1}$
and the sides of the cavity in $m$.\label{fig:30kS}}

\end{figure}

\section{Conclusion}

Numerical solutions of 2-D time dependent incompressible flow at high
Reynolds numbers in a driven cavity are presented. The numerical equations
are solved computationally using the numerical method above. For the
Reynolds numbers studied, up to $30\,000$, the solution converges
to a stationary state when using a very fine grid.

Based on the papers of Erturk et. al. \cite{Erturk05}, Erturk \cite{Erturk08}
and in this study, we conclude that in order to obtain a steady solution
for the driven cavity flow, a very fine grid mesh is necessary when
high Reynolds numbers are considered and also at high Reynolds numbers
when a coarse grid mesh is used then the solution oscillates. This
happens because a coarse mesh is not able to include the very small
vortices at the corners.

According to Erturk \cite{Erturk08}, the studies that presented unsteady
solutions of driven cavity flow using Direct Numerical Simulations
(DNS) (\cite{Auteri02,Cazemier98,Goyon96,Liffman96,Peng03,Poliashenko95,Tiesinga02,Wan02})
have experienced the same type of numerical oscillations because they
have used a coarse mesh. With this study we agreed with Erturk \cite{Erturk08}.
In all of the Direct Numerical Simulation studies on the driven cavity
flow found in the literature (\cite{Auteri02,Cazemier98,Goyon96,Liffman96,Peng03,Poliashenko95,Tiesinga02,Wan02}),
the maximum number of grid points used is less than $300\times300$.
Therefore, the periodic solutions found in \cite{Auteri02,Cazemier98,Goyon96,Liffman96,Peng03,Poliashenko95,Tiesinga02,Wan02}
are similar to the false periodic solutions observed by Erturk et.
al. \cite{Erturk05} and Erturk \cite{Erturk08} when a coarse grid
mesh is used.

In short, if a sufficiently fine grid mesh is used, a Direct Numerical
Simulation algorithm or other algorithm would give the same steady
results obtained by Erturk et. al. \cite{Erturk05}, Erturk \cite{Erturk08}
and in this study.

\section*{Acknowledgments}

This study was possible due to the computer cluster funded by FCT
grants PDCT/FP/63923/2005 and POCI/FP/81933/2007.

\end{document}